# Enhancing digital core image resolution using optimal upscaling algorithm: with application to paired SEM images

**Shaohua You[1], Shuqi Sun[1], Zhengting Yan[1], Qinzhuo Liao[1]\*, Huiying Tang[2], Lianhe Sun[1] and Gensheng Li[1]**

[1]National Key Laboratory of Petroleum Resources and Engineering, China University of Petroleum (Beijing), Beijing, 102249, China
[2]National Key Laboratory of Oil and Gas Reservoir Geology and Exploitation, Southwest Petroleum University, Chengdu, Sichuan 610500, China

\*Corresponding author. Email: liaoqz@cup.edu.cn.

**Summary**

The porous media community extensively utilizes digital rock images for core analysis. High-resolution digital rock images that possess sufficient quality are essential but often challenging to acquire. Super-resolution (SR) approaches enhance the resolution of digital rock images and provide improved visualization of fine features and structures, aiding in the analysis and interpretation of rock properties, such as pore connectivity and mineral distribution. However, there is a current shortage of real paired microscopic images for super-resolution training. In this study, we used two types of Scanning Electron Microscopes (SEM) to obtain the images of shale samples in five regions, with 1X, 2X, 4X, 8X and 16X magnifications. We used these real scanned paired images as a reference to select the optimal method of image generation and validated it using Enhanced Deep Super Resolution (EDSR) and Very Deep Super Resolution (VDSR) methods. Our experiments show that the bilinear algorithm is more suitable than the commonly used bicubic method, for establishing low-resolution datasets in the SR approaches, which is partially attributed to the mechanism of Scanning Electron Microscopes (SEM).

**Introduction**

Representative Elementary Volume (REV) properties are determined from 2D or 3D images obtained through CT or SEM scans. These images help assess the mechanical properties and permeability of geological porous media (Flannery et al., 1987; Fang et al., 2023; Kang et al., 2022; Yu et al., 2024). In the analysis of geological porous media, it is necessary to define a distinct Representative Elementary Volume (REV) for each mechanical and permeability property. The size and location of these REVs vary based on the specific property being analyzed, reflecting the diverse nature of geological formations. Compared to conventional laboratory core property measurements, the analysis of REV properties is faster and allows for numerical experiments on various properties of the same rock sample, including pore structure, mineral composition, and permeability (Krakowska et al., 2016; Starnoni et al., 2017; Chaves and Moreno, 2021; Liao et al., 2022). The computational accuracy of REV properties in geological porous media depends on the scanning resolution and image quality (Yang et al., 2021). However, due to constraints such as scanning costs and equipment limitations, it is often not possible to obtain nanoscale images of core samples, which leads to the omission of nanoscale pores and throats in numerical simulations. Additionally, environmental factors during the scanning process can also affect the quality of the rock images.

Super-resolution (SR) algorithms can be used to enhance image quality by restoring low-resolution images to high-resolution. In recent years, with the widespread application of Micro-CT, Nano-CT, and Scanning Electron Microscopes (SEM) in the field of petroleum engineering, digital rock technology has rapidly developed, and image super-resolution algorithms are increasingly applied in this field (Blunt et al., 2013; Xing et al., 2023; Yuan et al., 2023; Liang et al., 2024). Traditional image super-resolution methods based on prediction use algorithms such as bilinear, bicubic, or Lanczos, utilizing prior knowledge and improving edge detection to enhance image quality. Freeman et al. (2002) achieved improvements in image quality through interpolation, image rendering, and texture mapping. Inspired by manifold learning methods, Chang et al. (2004) employed a locally linear embedding (LLE) approach to reconstruct low-resolution images into high-resolution images, enhancing the compatibility and smoothness between high-resolution image patches. Li et al. (2017) proposed a 3D super-resolution algorithm based on sparse representation and combined it with a BM4D with a Laplacian filter for feature extraction, enhancing the resolution of reservoir rock CT images. This algorithm overcomes the inherent limitations of CT scan images and expands the field of view of high-resolution images. Although traditional SR methods are less computationally intensive, they do not adequately learn the detail differences between high and low-resolution images.

Deep learning methods surpass traditional super-resolution (SR) techniques by learning complex patterns and texture information in image data, producing clearer and more real high-resolution images. The application of deep learning methods in image super-resolution primarily relies on Convolutional Neural Network (CNN) and Generative Adversarial Network (GAN). CNN-based image super-resolution methods typically consist of three main convolutional layers: a layer to extract image features, a layer to map these features, and a layer for reconstructing the high-resolution output. Due to their stability, computational efficiency, and robust feature extraction capabilities, these methods have been widely adopted in the field of image super-resolution. Examples include the Super Resolution Convolutional Neural Network (SRCNN) (Lv et al., 2019), Very Deep Super Resolution (VDSR) (Kim et al., 2016), Enhanced Deep Super Resolution (EDSR) (Bizhani et al., 2022), Wide Activation Deep Super Resolution (WDSR) (Yu et al., 2018), and Fast Super Resolution Convolutional Neural Networks (FSRCNN) (Bai et al., 2020).Image super-resolution methods based on Generative Adversarial Networks (GANs) introduce an adversarial training mechanism, which is structured around a generator and a discriminator (Wang et al., 2018). The generator is responsible for producing high-resolution images, while the discriminator differentiates between the generated images and actual high-resolution images. Compared to CNNs, images generated by GANs have higher perceptual quality and stronger generalization capabilities (Ledig et al., 2016). Both methods have been broadly applied in the field of porous media image super-resolution (Kim et al., 2016; Shan et al., 2022; Yang et al., 2023). However, due to the high scanning costs of CT and SEM, image super-resolution methods in the porous media field typically involve generating synthetic datasets for training and evaluation. Low-resolution images are often created by downscaling real high-resolution images using the bicubic method. This approach significantly simplifies the super-resolution process but results in substantial differences between the synthetic low-resolution images and actual low-resolution images.

In recent years, researchers have begun to establish authentic super-resolution datasets for the study of paired low-resolution (LR) and high-resolution (HR) image super-resolution algorithms. Cai et al. (2019) adjusted the focal length of digital cameras to create an authentic super-resolution dataset and proposed a Laplacian pyramid-based kernel prediction network (LP-KPN) that effectively restores high-resolution images. Zhang et al. (2019) trained a deep network with a bilateral loss using an authentic super-resolution dataset and demonstrated its superior performance with 4x and 8x image training. Wei et al. (2020) established a new super-resolution (SR) benchmark with real high-resolution image upscaling processes, reducing the limitations of traditional interpolation methods. In the field of porous media, the application of authentic super-resolution datasets is less common. Chen et al. (2020) used a CycleGAN approach to achieve super-resolution of unpaired rock CT images. Zhao et al. (2023) used the same dataset as Chen and, by grouping the training dataset according to the size ratio of pore throats and voxels, reconstructed images with better clarity and contrast. Although using authentic super-resolution datasets can provide more reliable SR analysis results, obtaining real different resolution core images with CT or SEM scanning, compared to digital camera photography, involves significantly higher costs and more challenging image pairing.

In this study, we obtained real paired SEM microscopic images at different resolutions through scanning. We used a Hitachi SU8010 Field-Emission Scanning Electron Microscope (FE-SEM) and a Quanta 200F Field-Emission Environmental Scanning Electron Microscope (FEE-SEM) to capture grayscale images of five types of shale from different regions, taking five photographs of each region at magnifications of 1X, 2X, 4X, 8X and 16X. After pairing images of different resolutions, we used six methods—nearest, box, bilinear, bicubic, Lanczos2, and Lanczos3—to upscale real high-resolution images, and we quantitatively compared the differences between the generated low-resolution images and the actual low-resolution images using PSNR and SSIM. Similarly, we downscaled real low-resolution images to compare the generated high-resolution images with actual high-resolution images. Bilinear was selected as the optimal method for both upscaling and downscaling. Finally, we validated our study using two image super-resolution methods, VDSR and EDSR, which confirmed the effectiveness of our research.

## Methodology

### Binary Image of SEM

In this study, to demonstrate the generalizability of our research, we acquired original images of shale samples from five regions (i.e., Gulong, Jiyang, Jimusar, Yanchang, and Changqing) using two types of scanning electron microscopes (SEMs): Quanta 200F Field-Emission Environmental Scanning Electron Microscope (FEE-SEM) and Hitachi SU8010 Field-Emission Scanning Electron Microscope (FE-SEM). The Quanta 200F FEE-SEM, manufactured by Thermo Fisher Scientific from the USA, features a field emission gun that supports high vacuum, low vacuum, and environmental SEM modes, making it highly versatile for studying various sample types under different conditions. The Hitachi SU8010 FE-SEM, made by Hitachi High-Tech Corporation in Japan, is equipped with a secondary electron detector and an energy-dispersive X-ray (EDX) detector, specifically the Bruker EDX Detector and EDX XFLASH Detector, which enhance its analytical capabilities for detailed compositional analysis. The FEE-SEM was used to scan shale samples from all five regions, capturing grayscale images in five different areas of each sample at magnifications of 1X, 2X, 4X, 8X and 16X. The voxel lengths for these magnifications were 9.1 nm, 18.2 nm, 36.4 nm, 72.8 nm and 145.6 nm, respectively. The FE-SEM was utilized specifically for scanning shale samples from the Gulong and Jiyang regions for comparative purposes, also at five magnifications. The corresponding voxel lengths for these magnifications were 12.4 nm, 24.8 nm, 49.6 nm, 99.2 nm, and 198.4 nm (Fig. 1).

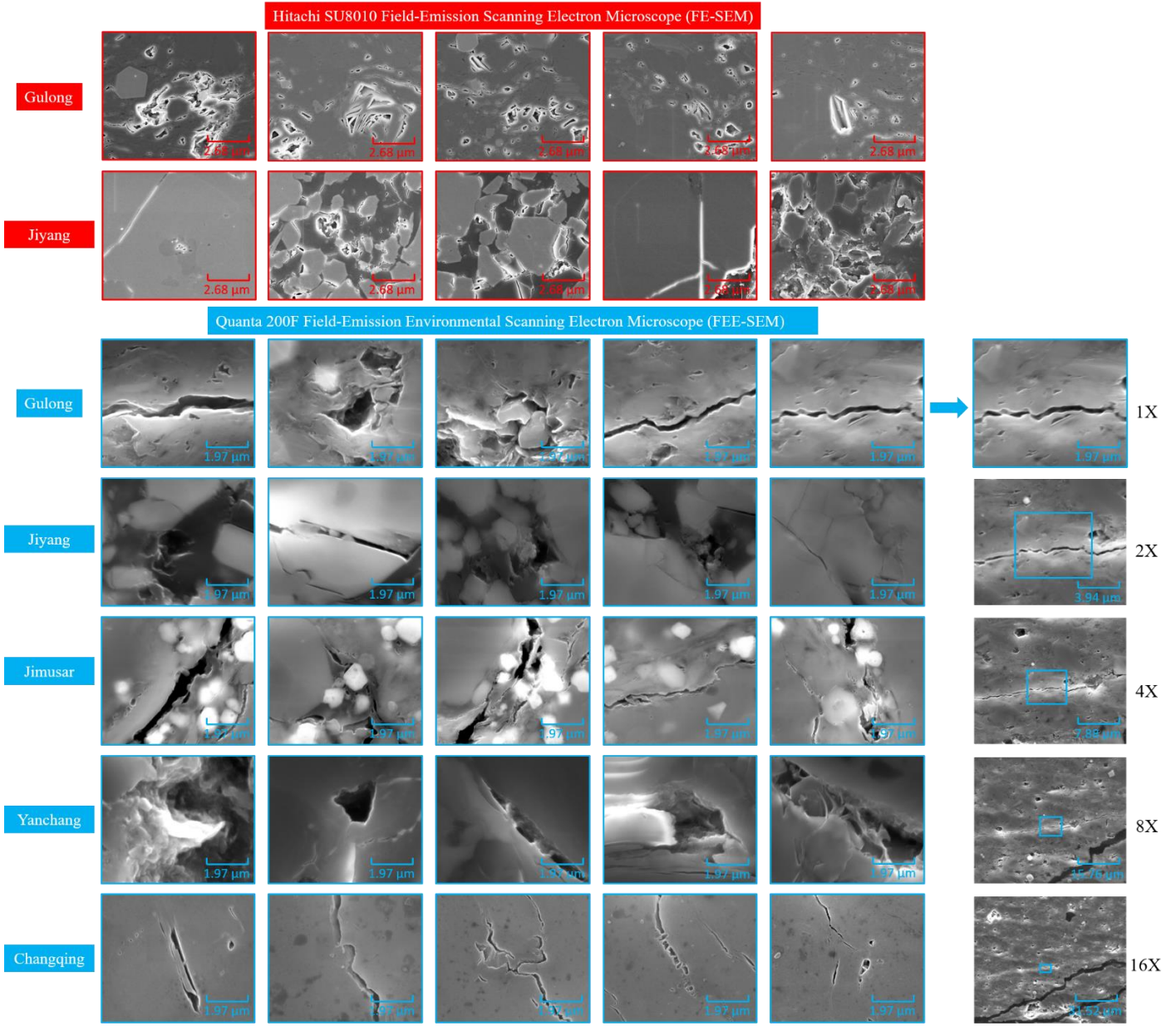

**Fig. 1—The original grayscale images of SEM scanning of shale samples in different regions.**

**Image Generation Method**

After obtaining grayscale images at various voxel lengths through scanning, image pairing was performed to facilitate subsequent comparisons of image upscaling and downscaling methods. Initially, interpolation algorithms such as nearest neighbor, box, bilinear, Lanczos2, and Lanczos3 were employed to upscale high-resolution images and extract their intensity values, all yielding equivalent results in image matching. Subsequently, we utilized the method of normalized cross-correlation for pairing images of different voxel lengths (Lewis, 1995):

$$\text{NCC}(x, y) = \frac{\sum_{x',y'} \left[ T(x', y') \cdot I(x+x', y+y') \right]}{\sqrt{\sum_{x',y'} T(x', y')^2 \cdot \sum_{x',y'} I(x+x', y+y')^2}} \tag{1}$$

where $T$ represents the template image, and $I$ is the image being searched, with $x'$ and $y'$ as the coordinates within the template. We used the 1X grayscale image as the template to pair with the 2X, 4X, 8X and 16X images.

Due to environmental influences during SEM imaging, grayscale values of images at different voxel lengths exhibited varying distributions. Therefore, it was necessary to normalize the paired images to mitigate the impact of environmental factors on subsequent computations. We employed the Z-score normalization method, the approach involves calculating the mean and standard deviation of the image data (Iglewicz and Hoaglin, 1993):

$$Z = \frac{X - \mu}{\sigma} \quad (2)$$

where Z is the Z-score of the image data, X is the original value from the image data, $\mu$ is the mean of all image data values, and $\sigma$ is the standard deviation of the image data. This processing step ensures that the normalized images have zero mean and unit standard deviation, thereby eliminating differences between images caused by environmental influences.

In this study, we tested six different image upscaling and downscaling algorithms, including the commonly used bicubic method in the super-resolution (SR) domain, as well as nearest, box, bilinear, Lanczos2, and Lanczos3 algorithms. The nearest and box algorithms are relatively simple. The nearest method directly assigns to each new pixel a value from the nearest original pixel during upscaling and downscaling (Hu et al. 2006):

$$f(x, y) = I(round(x), round(y)) \quad (3)$$

where $I$ is the input image, and $x$, $y$ are the coordinates in the target image, where the function round rounds these coordinates to the nearest integer, representing the closest pixel position in the source image.

The box method determines each pixel in the target image by calculating the average value of the pixels in the corresponding area of the source image (Hu et al. 2006):

$$f(x, y) = \frac{1}{N} \sum_{i=1}^{N} I(x_i, y_i) \quad (4)$$

where $N$ is the number of pixels in the area, and ($x_i$, $y_i$) are the pixel coordinates within that area, typically a small window.

The bilinear method, more complex than the nearest and box methods, yields smoother image edges. It calculates the weighted average of the four nearest pixels to the target pixel based on their distance (Parks et al. 1987):

$$f(x, y) = \frac{1}{(x_2 - x_1)(y_2 - y_1)} \sum_{i=1}^{2} \sum_{j=1}^{2} I(x_i, y_j) \cdot (x_2 - x)^{1-i} \cdot (y_2 - y)^{1-j} \quad (5)$$

where, $I(x_i, y_i)$ represents the four pixels closest to the target pixel ($x$, $y$). This formula first performs linear interpolation in one direction, followed by a second linear interpolation on the result.

The bicubic, a common algorithm for image upscaling and downscaling in image super-resolution, provides smoother interpolation than bilinear interpolation, maintains clearer edges, and better preserves details when enlarging images. It estimates new pixel values based on the sixteen pixels surrounding the target pixel using a cubic polynomial function (Hu et al. 2006):

$$f(x, y) = \sum_{i=-1}^{2} \sum_{j=-1}^{2} I(x_i, y_j) \cdot h(x - x_i) \cdot h(y - y_j) \quad (6)$$

where $h(t)$ is the cubic Hermite interpolation kernel, commonly defined with:

$$h(t) = \begin{cases} (a+2)|t|^3 - (a+3)|t|^2 + 1 & \text{if } |t| \leq 1 \\ a|t|^3 - 5a|t|^2 + 8a|t| - 4a & \text{if } 1 < |t| \leq 2 \\ 0 & \text{otherwise} \end{cases} \quad (7)$$

And the adjustable parameter $a$ is typically set at –0.5.

Lanczos2 and Lanczos3 algorithms, based on windowed *sinc* (sine/sampling function) interpolation, provide high-quality image reconstruction, maintaining excellent detail and edge clarity. The primary difference between Lanczos2 and Lanczos3 lies in the range of adjacent pixels used; Lanczos3 uses a broader range than Lanczos2, thus achieving higher image quality but at a greater computational complexity (Duchon, 1979):

$$f(x, y) = \sum_{i=-a+1}^{a} \sum_{j=-a+1}^{a} I(x_i, y_j) \cdot sinc(x - x_i) \cdot sinc(y - y_j) \cdot sinc\left(\frac{x - x_i}{a}\right) \cdot sinc\left(\frac{y - y_j}{a}\right) \quad (8)$$

where $sinc(x) = \frac{\sin(\pi x)}{\pi x}$ represents the standard *sinc* function, and a is the size of the Lanczos window, with $a=2$ for Lanczos2 and $a=3$ for Lanczos3.

Using these six image generation methods, we generated six low-resolution images from real high-resolution images through upscaling, and six high-resolution images from real low-resolution images through downscaling. We then selected the most accurate upscaling and downscaling algorithms by comparing the generated images with the actual SEM images.

**Image Super Resolution**

In the field of digital rock image super-resolution, there are typically two forms of image training sets: one uses grayscale images directly from SEM and CT scans, while the other employs binary images resulting from threshold segmentation of the grayscale images. Due to the minor differences between grayscale images of different resolutions, the first training method is usually applied to digital rock image training sets that contain a substantial volume of image data. In our study, due to the constraints on scanning costs, only five grayscale images were scanned for each region's shale sample. Moreover, the variation between images from the same region was significant, precluding the simultaneous training of residual networks. Consequently, we opted for the second type of training set, using binary images for training.

Threshold segmentation is a crucial step in the field of digital rock core analysis. Binary images resulting from threshold segmentation are utilized for numerical simulation of flow parameters such as porosity, permeability, and tortuosity. Compared to grayscale images, binary images at different resolutions exhibit greater differences, making them preferable for training in digital rock image super-resolution. In our study, we have paired and normalized grayscale images of different magnifications, enabling us to use the same grayscale value as the threshold for segmenting images at different magnifications. Subsequently, the generated binary images were used as the training set for training VDSR and EDSR networks.

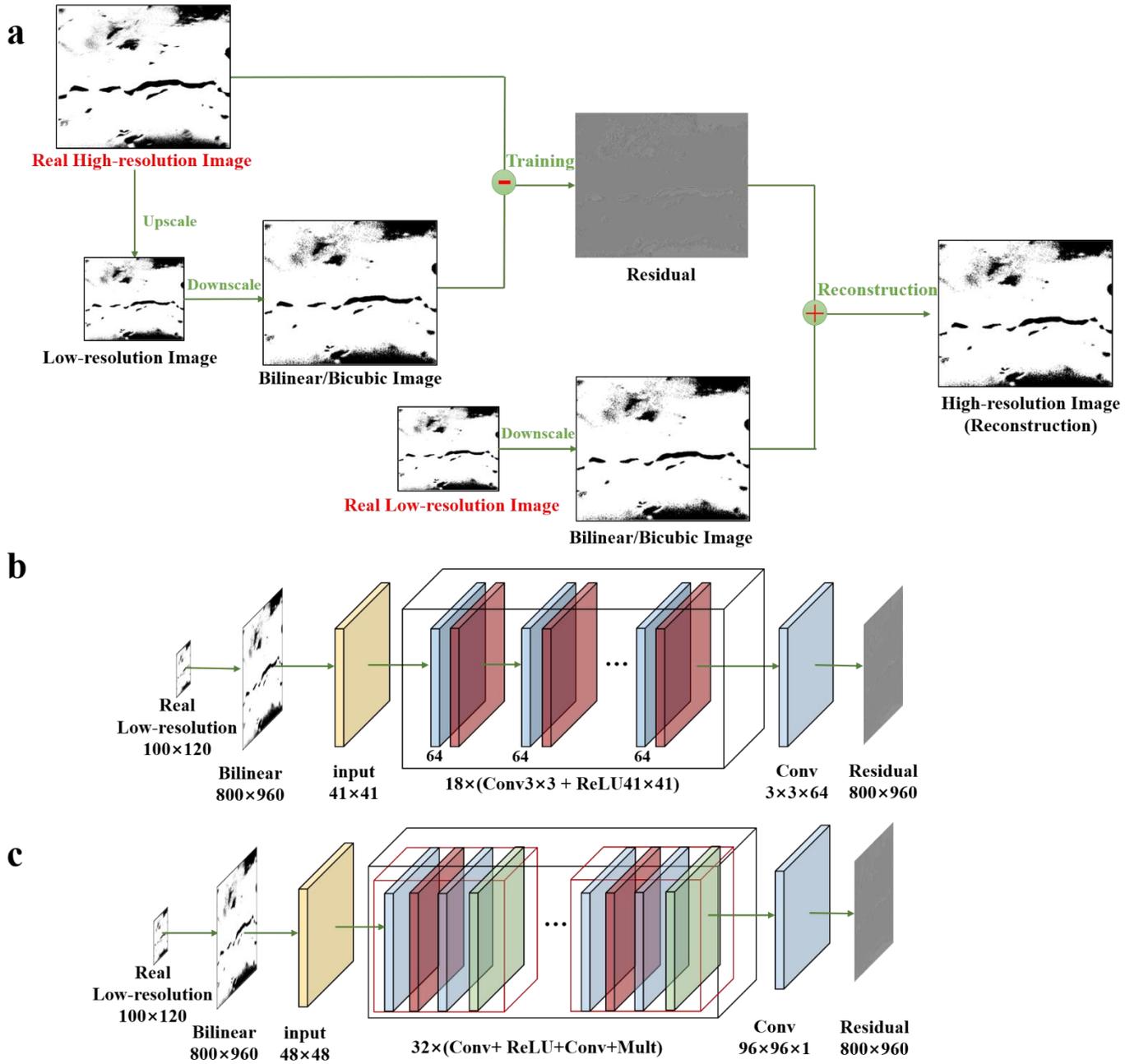

**Fig. 2—Illustration of (a) the process of super resolution, (b) the structure of VDSR and (c) the structure of EDSR.**

VDSR and EDSR, based on convolutional neural networks, train on the residuals between high-resolution and low-resolution images to effectively reconstruct high-resolution images. This method consists of three steps: The first step involves training the convolutional neural network to learn the differences in detail between high-resolution and low-resolution images. The second step creates residual images by feeding the low-resolution images into the trained network, yielding residuals corresponding to those images. The third step is image reconstruction, where the residual images are added back to the low-resolution images to reconstruct new high-resolution images.

Fig. 2a displays the flowchart of the image super-resolution process in this study. We first upscale real high-resolution images to generate low-resolution images, then downscale these images to restore the original size. Subsequently, we use the VDSR or EDSR network to learn the detailed features between the real high-resolution images and the generated low-resolution images. The error function during training is defined as the difference between the high-resolution and low-resolution images. As the real high-resolution images are binarized, and the upscaled low-resolution images are continuous, the residual images obtained through training are also continuous. After completing the training of the VDSR or EDSR network, we apply the trained residual network to reconstruct high-resolution images using real scanned low-resolution images. Since the training set employs binarized images, while the residuals generated through training are continuous (ranging from 0 to 1), it is necessary to apply a threshold of 0.5 to the reconstructed images for subsequent threshold segmentation. This process validates the selected image upscale and downscale methods.

Fig. 2b presents the detailed structure of the VDSR network. The input layer has a window size of 41×41, used to extract local features from high-resolution and low-resolution images. The middle layers include 18 alternating convolutional and rectified linear unit (ReLU) layers, used for performing non-linear mapping to learn the differences between high-resolution and low-resolution images. The final layer is a convolutional layer of size 3×3×64, used for constructing the residual image. Overall, this process is a form of image synthesis, mapping the features of high-resolution and low-resolution images back into pixel space to generate residual images (Dong et al., 2014; Vb et al., 2020).

Specifically, the input layer serves as the first feature extraction layer (41×41×1) to identify and extract the differences in detail between high-resolution and low-resolution images (Vb et al., 2020):

$$F_1(Y) = \max(0, W_1 g Y + B_1) \tag{9}$$

where $Y$ represents the real low-resolution image, $W_1$ and $B_1$ are the parameters of the filter (convolution kernel), representing the weight and bias of the filter, and the max operation corresponds to the ReLU activation function.

The extracted detail features are then filtered and activated through the ReLU layer, a step that is iterated 18 times to enhance the learning capacity of the network (Vb et al., 2020):

$$F_2(Y) = \max(0, W_2 g F_1(X) + B_2) \tag{10}$$

where $W_2$ and $B_2$ represent the weight and bias of the filter.

Finally, a convolutional layer is used to construct the residual image. Because the image size reduces after the initial convolutions, this step uses a deconvolutional approach to restore the image size:

$$F(Y) = W_3 g F_2(Y) + B_3 \tag{11}$$

where $W_3$ and $B_3$ are still the parameters of the filter (convolution kernel).

The detailed structure of the EDSR network, as shown in Fig. 2c, is similar to that of the VDSR network, but it differs mainly in the size of the windows and the number of parameters (Lim et al., 2017). In this EDSR network architecture, the input layer and the middle layers utilize a window size of 48×48, while the final convolutional layer uses a window size of 96×96. Additionally, Lim et al. (2017) introduced a multi-scale model within the network architecture, which can effectively reconstruct high-resolution images at various scales. Compared to the VDSR network structure, EDSR has a greater number of parameters and iterations, allowing for better learning of the detailed features of pores and throat edges in the binary rock core images.

Through the EDSR and VDSR networks, we learn the differences in detail between real high-resolution images and generated low-resolution images. After training, the residual networks are applied to real low-resolution images. By quantitatively comparing the percentage improvement in the quality of real low-resolution images before and after super-resolution (SR), we can validate the optimized upscaling and downscaling methods.

**Image Quality Evaluation Index**

Peak Signal-to-Noise Ratio (PSNR) and Structural Similarity Index (SSIM) are two commonly used image quality assessment metrics in the field of image super-resolution. PSNR is a measure of image distortion and is typically used to measure the similarity between an original image and one that has been compressed or processed in other ways. It calculates the peak signal-to-noise ratio between the original and processed images. The mean square error (MSE), which is a component of PSNR, is calculated as follows (Wang et al., 2004):

$$MSE = \frac{1}{mn} \sum_{i=0}^{m-1} \sum_{j=0}^{n-1} \|I(i,j) - K(i,j)\|^2 \tag{12}$$

where $m$ and $n$ represent the size of the image m×n, and $I$ and $K$ represent the two images being compared. The peak signal-to-noise ratio is defined as (Wang et al., 2004):

$$PSNR = 10 \cdot \log_{10}(\frac{MAX_I^2}{MSE}) = 20 \cdot \log_{10}(\frac{MAX_I}{\sqrt{MSE}}) \tag{13}$$

where $MAX_I$ is the maximum possible pixel value of the image. The higher PSNR value indicates better image quality and closer resemblance to the original high-resolution image.

SSIM, on the other hand, is an image quality metric that considers the perceptual characteristics of human vision, assessing image quality by simulating the human visual system. It encompasses three dimensions: luminance, contrast, and structure. Unlike PSNR, SSIM is more closely related to human perception because it considers the structural information of the image. The formula for SSIM is:

$$SSIM(x, y) = \frac{(2\mu_x\mu_y + C_1)(2\sigma_{xy} + C_2)}{(\mu_x^2 + \mu_y^2 + C_1)(\sigma_x^2 + \sigma_y^2 + C_2)} \tag{14}$$

where $\mu$ represents Luminance, $\sigma$ represents Contrast, and $C$ represents Structure. The SSIM value ranges from 0 to 1, with a value closer to 1 indicating higher similarity between the processed and original images.

In applying various image upscaling methods to generate low-resolution images, we used PSNR and SSIM to quantitatively evaluate the differences between the generated low-resolution images and high-resolution images, selecting the best image upscaling method. Similarly, we also identified the best image downscaling method. The results were validated using the EDSR and VDSR networks.

## Result

### Image Upscaling

By comparing the PSNR and SSIM of the six low-resolution images generated by various methods (nearest, box, bilinear, bicubic, Lanczos2, and Lanczos3) with the real low-resolution images, we can identify the most accurate image upscaling method. Using the example of fracture images of shale samples from the Gulong region scanned by the Quanta 200F Field-Emission Environmental Scanning Electron Microscope (FEE-SEM), we demonstrated the comparison of low-resolution images at 2X, 4X, 8X and 16X generated by different image upscaling methods with the real low-resolution images (Fig. 3). It is evident that when real high-resolution images are upscaled by factors of 2 and 4 using different interpolation algorithms, the generated low-resolution images show little difference from the real scanned low-resolution images. We have also conducted a frequency domain analysis of grayscale images. For further details on the subtle differences in various frequency domain analyses, refer to Fig. S-1 in the Supplementary Materials. However, as the real high-resolution images are upscaled by factors of 8 and 16, the differences between the images generated by various upscaling methods become more noticeable. Visually, images produced using the nearest and box algorithms show significant differences from the real low-resolution images, especially in the areas around pores and fractures.

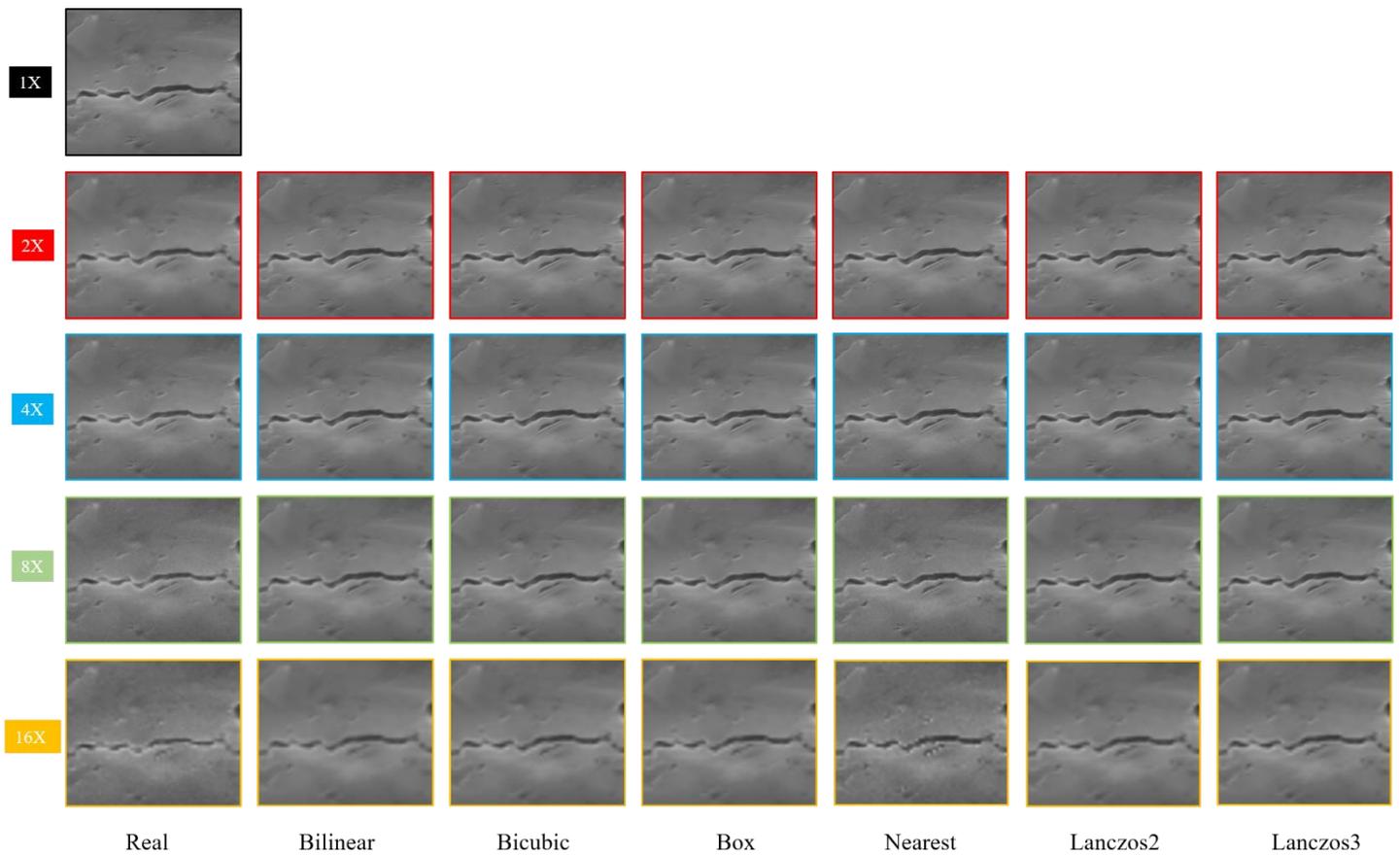

**Fig. 3—Comparison between the low-resolution images generated by upscaling of six methods and the real scanned low-resolution images.**

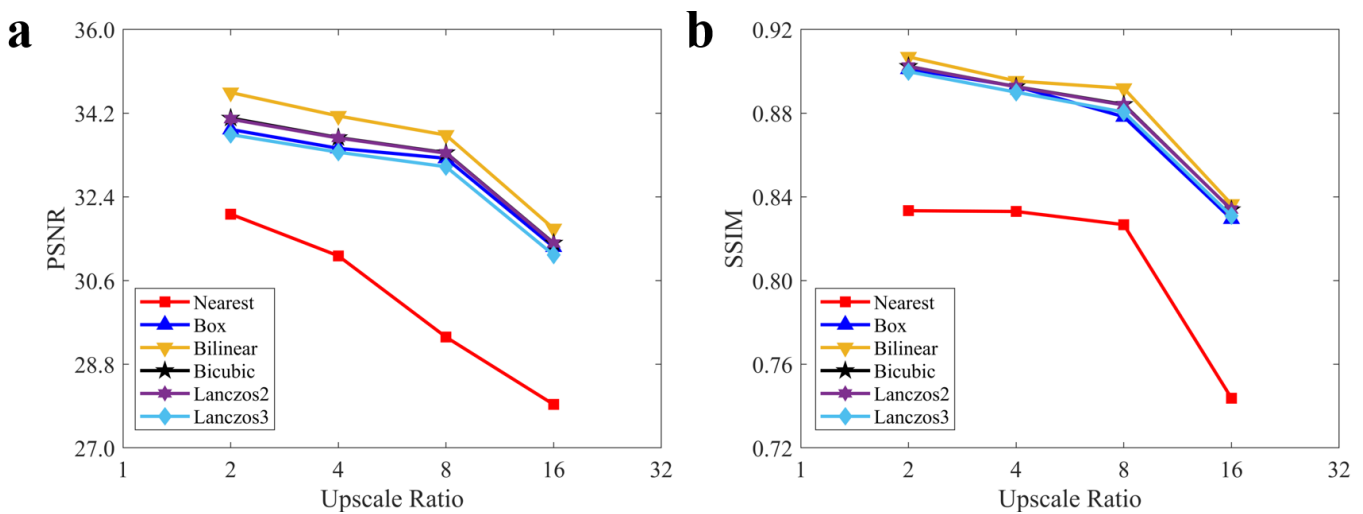

**Fig. 4— (a) PSNR and (b) SSIM values of the low-resolution images generated by upscaling of different methods compared to the real low-resolution images.**

Fig. 4a displays the PSNR values of low-resolution images generated using different upscaling methods compared to the real low-resolution images. The x-axis represents the upscale ratio of the real high-resolution images, and the y-axis shows the PSNR values. The line graph indicates that as the upscale ratio increases, the PSNR value decreases, indicating increased image distortion and decreased clarity. Comparing different upscaling methods, regardless of whether the upscale ratio is 2, 4, 8, or 16, images generated using the bilinear algorithm consistently show the highest PSNR values. The box, bicubic, Lanczos2, and Lanczos3 algorithms have closely competing PSNR values, while the nearest algorithm produces the lowest PSNR values. This indicates that among the six upscaling methods, the Bilinear method generates low-resolution images that are closest to the real low-resolution images.

Fig. 4b shows the SSIM values of low-resolution images generated using different upscaling methods as the upscale ratio changes. Similar to the trend in PSNR values, as the upscale ratio increases, the SSIM values between the generated and real low-resolution images decrease. Among the different upscaling methods across all four magnification levels, the bilinear algorithm consistently results in the highest SSIM values, indicating that from the dimensions of luminance, contrast, and structure, images generated using the Bilinear method are closest to the real low-resolution images. Therefore, through quantitative analysis of PSNR and SSIM metrics for low-resolution images generated through upscaling compared to real low-resolution images, the Bilinear method is identified as the optimal choice.

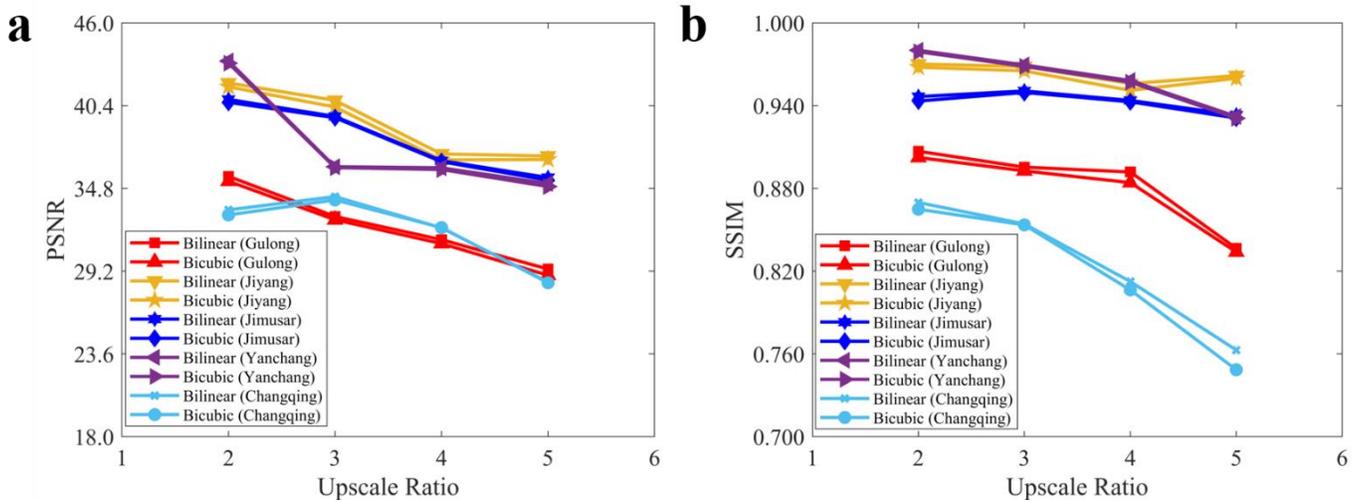

**Fig. 5—** (a) PSNR and (b) SSIM values of the low-resolution images generated by upscaling from different regions compared to the real low-resolution images.

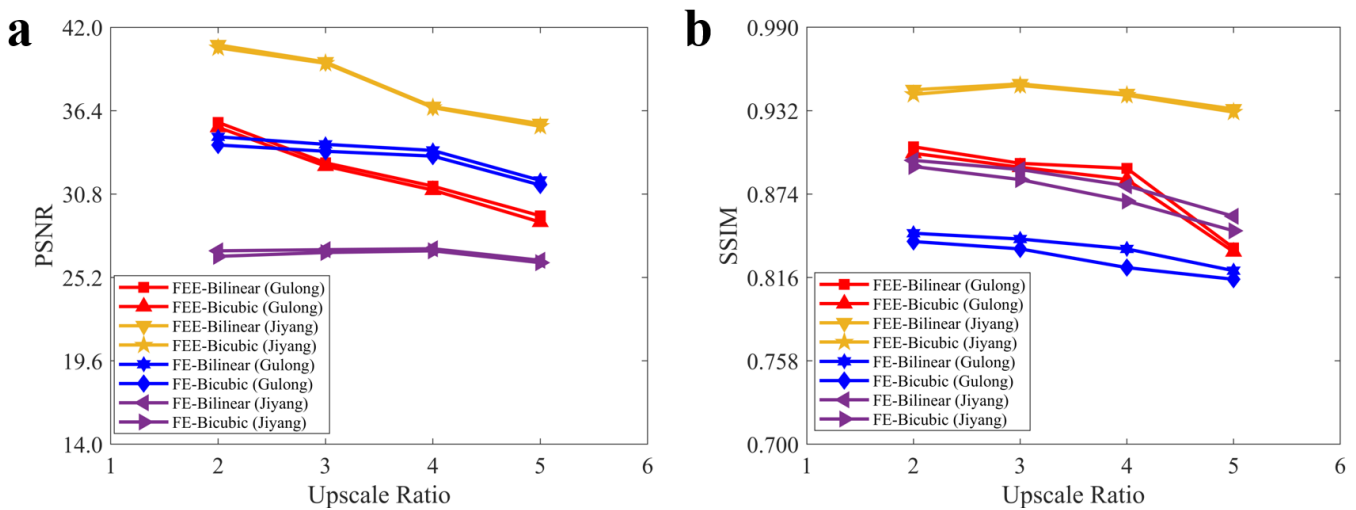

**Fig. 6—** (a) PSNR and (b) SSIM values of the upscaled low-resolution image and the real low-resolution image scanned by FE-SEM and FEE-SEM.

The common practice among researchers in the digital rock super-resolution field is to use the bicubic method. Hence, we compared SEM images from different regions and those taken using different SEMs to validate our conclusions. Figs. 5a and 5b show the PSNR and SSIM values for low-resolution images generated from rock samples in five regions and the corresponding real low-resolution images, all scanned by the FEE-SEM. The comparison of bilinear and bicubic methods across these regions generally follows the trend of decreasing PSNR and SSIM values with increasing upscale ratios. Despite potential interference from environmental factors, which may cause significant differences in PSNR and SSIM values across different magnifications for rock samples from different regions, images generated using the bilinear method consistently show higher PSNR and SSIM values compared to those generated using the bicubic method, regardless of the region. This indicates that the bilinear method generates low-resolution images that are closer to the real low-resolution images.

To prove that this conclusion is not solely applicable to FEE-SEM grayscale images, we conducted a comparative validation using grayscale images scanned by FE-SEM (Figs. 6a and 6b). FE-SEM only scanned shale samples from the Gulong and Jiyang regions. For the images scanned using FE-SEM, the bilinear method produced images with higher PSNR and SSIM values. The patterns observed confirm the conclusion, further demonstrating its validity.

**Image Downscaling**

In the research on digital rock image super-resolution, paired images are used for model training. As low-resolution images are smaller in size than high-resolution images, it is necessary to downscale low-resolution images to unify the sizes of images at different resolutions for super-resolution training. Therefore, it is crucial to identify the most accurate image downscale method that closely matches the real paired images.

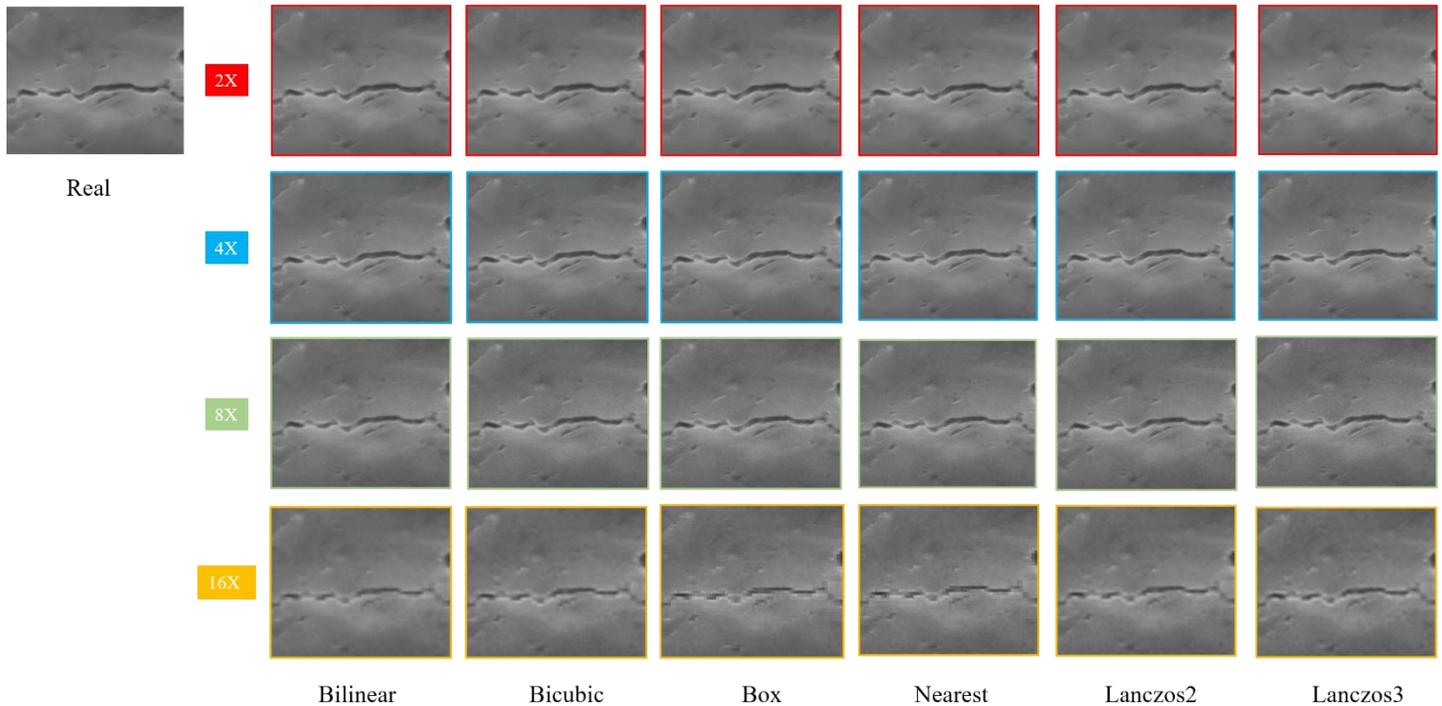

**Fig. 7—Comparison between the low-resolution images generated by downscaling of six methods and the real scanned high-resolution images.**

We employed six methods—nearest, box, bilinear, bicubic, Lanczos2, and Lanczos3—to downscale low-resolution images by factors of 2, 4, 8, and 16, and compared these downscaled images with the actual high-resolution images to determine the best method. Fig. 7 shows the results of images generated by different methods. Images downscaled from real low-resolution images at 2X, 4X, and 8X scales show minor differences from real high-resolution images, and the disparity among the methods is also small. However, at a 16X downscale factor, the differences between the methods become significantly noticeable. Images downscaled using nearest and box methods exhibit considerable distortion, especially around the edges of pores and fractures. Images produced by bilinear, bicubic, Lanczos2, and Lanczos3 methods are closer to the real high-resolution images, but visually it is challenging to discern the similarity between these images and the actual high-resolution images; hence, a quantitative analysis from image quality parameters is needed.

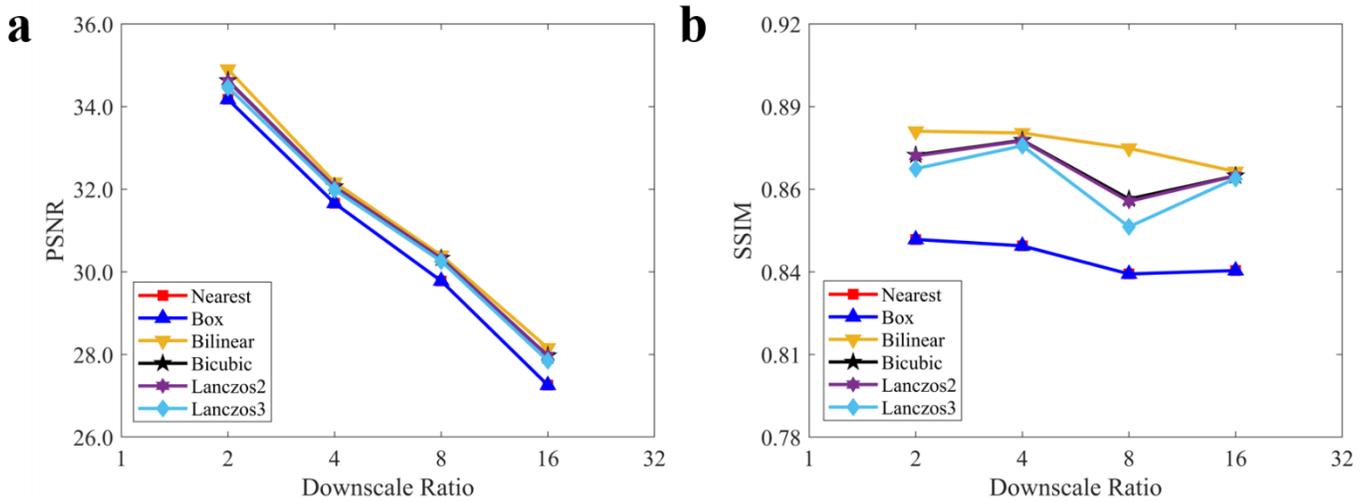

**Fig. 8—** (a) PSNR and (b) SSIM values of the low-resolution images generated by upscaling of different methods compared to the real low-resolution images.

Fig. 8 displays the PSNR and SSIM values for images resized using different methods compared to the actual high-resolution images. The x-axis represents the downscale ratio of the real low-resolution images, and the y-axis shows the PSNR and SSIM values. The graph indicates that as the image downscale ratio increases, the PSNR values decrease, implying greater image distortion. However, the pattern for SSIM values in Fig. 8b does not align with this, possibly due to inconsistencies in the scanning environment. Comparing the curves for different downscale methods, images downscaled using nearest and box methods have the lowest PSNR and SSIM values, suggesting that the principles behind these methods differ significantly from those of SEM scanning. From the perspective of peak signal-to-noise ratio, the images generated by bilinear, bicubic, Lanczos2, and Lanczos3 methods are similar, with the Bilinear method slightly outperforming the others. From the standpoint of Structural Similarity, images downscaled using the Bilinear method are closest to the real high-resolution images and markedly superior to the other methods. Similar to tests with image upscale methods, we compared rock sample images from different regions and images scanned by different SEM. Figs. S-2 and S-3 in the Supplementary Materials present the results for PSNR and SSIM. Therefore, by quantitatively analyzing the PSNR and SSIM values of images generated by these methods compared to real high-resolution images, it is evident that the bilinear method's principles most closely resemble those of SEM scanning.

Through qualitative comparisons and quantitative analyses of image quality parameters for images generated by the nearest, box, bilinear, bicubic, Lanczos2, and Lanczos3 methods for both upscaling and downscaling, we find that images generated using the Bilinear method are the closest match to the real paired images.

**Enhanced Deep Super Resolution (EDSR)**

In this study, we utilized the bilinear and bicubic methods to generate paired images for establishing a digital rock super-resolution dataset, which was then used for training convolutional neural networks. The trained residual networks were applied to real low-resolution images to reconstruct high-resolution images, thereby demonstrating the superiority of the bilinear method over the bicubic. Due to scanning cost limitations, we only scanned five sets of images from each rock sample region. However, the grayscale differences between images at different resolutions were minimal, especially for the 2X and 4X paired images, making it challenging for EDSR and VDSR to effectively learn the residual details between paired grayscale images under conditions of limited training samples. Therefore, we performed threshold segmentation on grayscale images to enhance the detail differences between images generated by the two methods (Fig. 9).

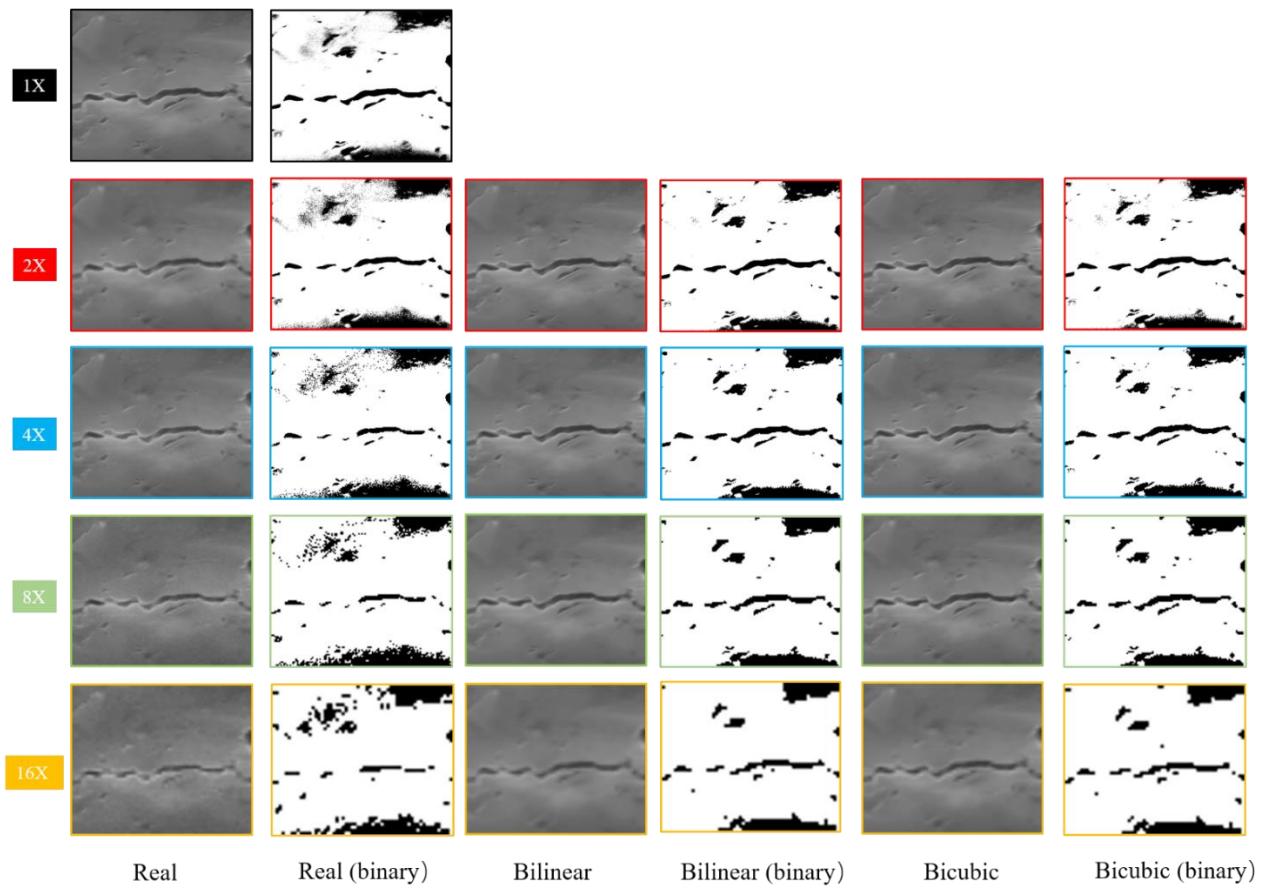

**Fig. 9—Comparison between grayscale images and binary images.**

We converted the paired images generated by the bilinear and bicubic methods into binary images, which served as the super-resolution training set for EDSR training. Due to the significant distortion in images upscaled by 16 times from real high-resolution images, EDSR struggled to learn the detail differences between these images and the actual high-resolution images. Consequently, we only displayed the image reconstruction results for the 2X, 4X and 8X paired binary images (Fig. 10). The images reconstructed using the EDSR network showed a noticeable improvement, especially in restoring small pores and micro-fractures, compared to the direct upscale generated images using bilinear and bicubic methods. The main objective of this study was to identify the image generation method that most closely resembles the principles of scanning electron microscope (SEM) scanning and to establish a SEM digital rock super-resolution training dataset based on this principle, rather than enhancing the performance of existing super-resolution methods. Thus, we needed to use PSNR and SSIM as image quality metrics to quantitatively analyze the paired training sets generated by the two methods.

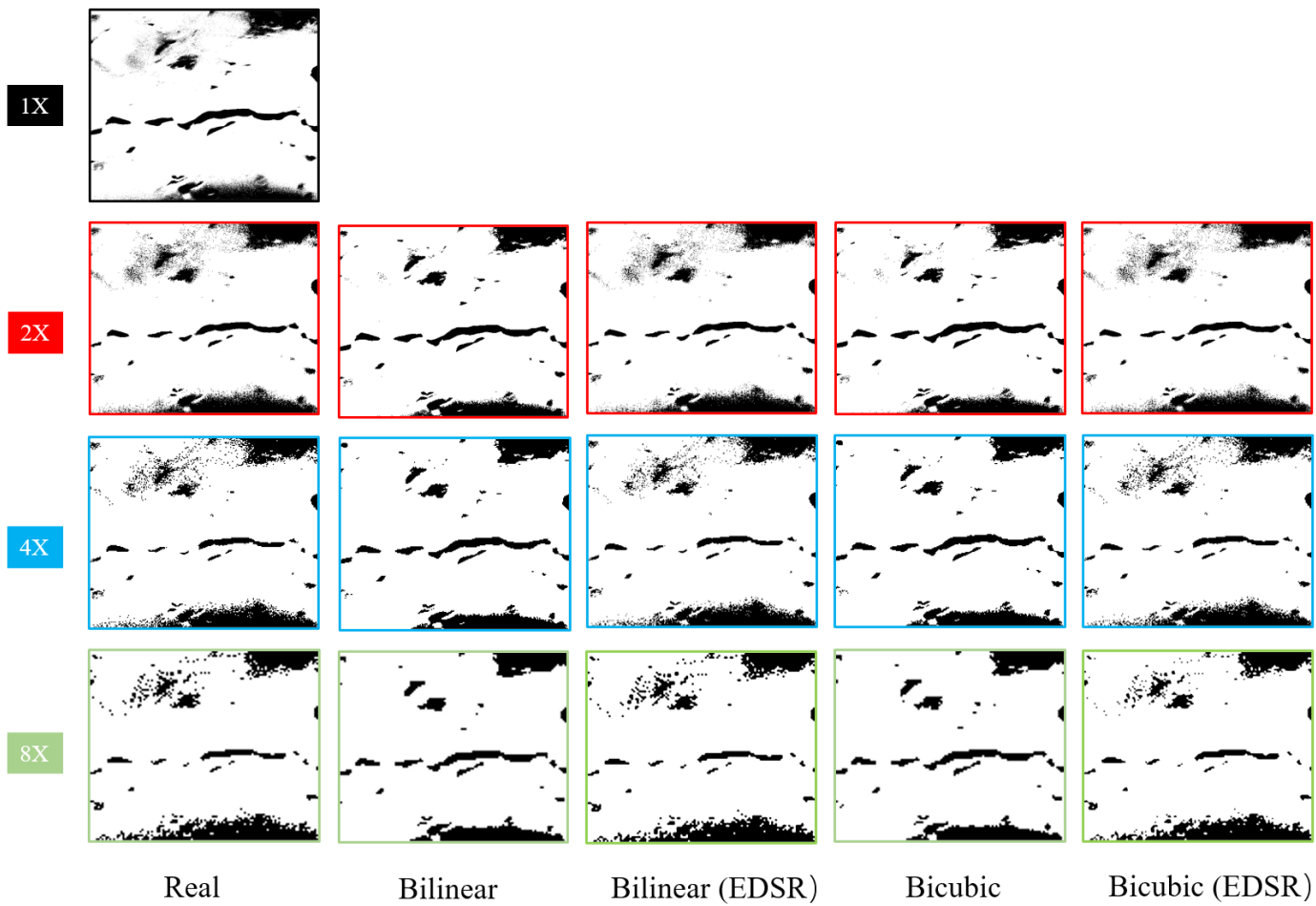

Fig. 10—High-resolution images reconstructed using residual networks trained with different EDSR training sets and real low-resolution images.

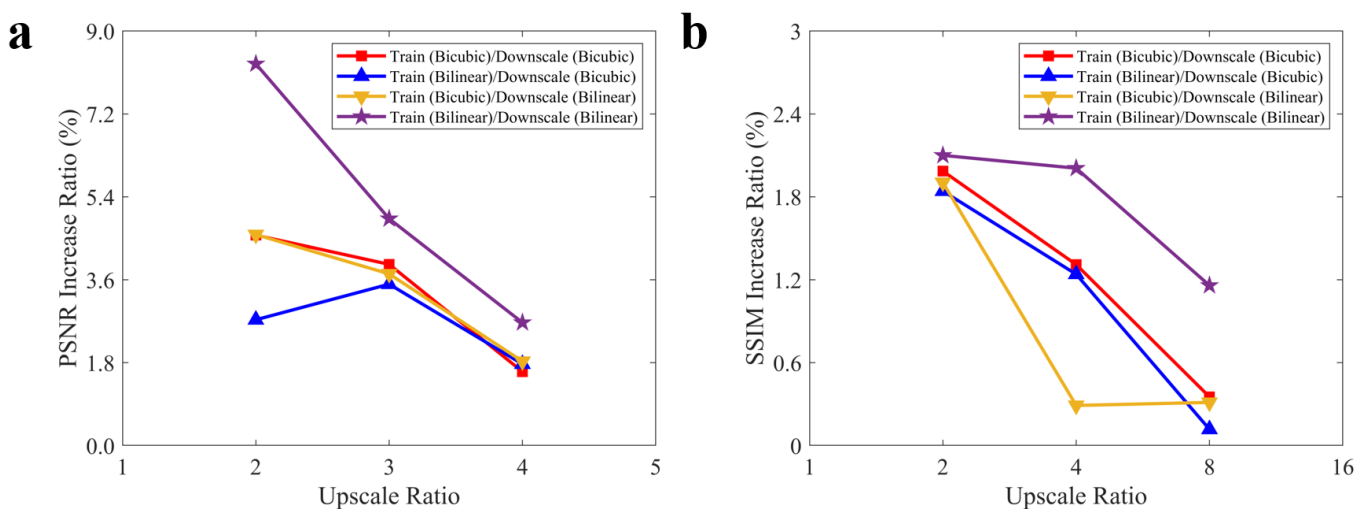

Fig. 11—Comparison of (a) PSNR increase rates and (b) SSIM increase ratio of real low-resolution images by different EDSR training sets and image downscale method.

Fig. 11 displayed the percentage improvement in PSNR and SSIM before and after using the residual network trained with two different datasets, generated by the bilinear and bicubic methods at various resolutions. During the image reconstruction process, we first upscaled the real low-resolution images to match the size of high-resolution images. In the resizing step, we used both bilinear and bicubic methods. As shown in Fig. 11a, when the same method was used for generating paired images and resizing (either bilinear or bicubic), the real low-resolution images showed the highest increase in PSNR after super-resolution. Comparing the two methods, using bilinear-generated paired images as the training set was significantly superior to the bicubic method. Since EDSR learns the detail differences between real high-resolution images and those generated by the bilinear method, and image reconstruction was performed using real low-resolution images, it suggests that from a peak signal-to-noise ratio perspective, bilinear-generated images are closer to real low-resolution images. Fig. 11b displayed the SSIM results, which followed the same pattern as Fig. 11a, showing that using the same method for both generating paired images and resizing during the image reconstruction process led to the highest increase in SSIM. This indicates that regardless of the method used to generate the super-resolution dataset, the same method should be used to resize images during the image reconstruction process. Compared between the two methods at 2X, 4X, and 8X magnifications, the training set generated by the bilinear method was superior to that generated by the bicubic method, demonstrating that from the perspectives of luminance, contrast, and structure, the bilinear algorithm-generated paired images are closer to real low-resolution images.

Fig. 12 showed the improvement rates in image quality metrics before and after super-resolution of real low-resolution images using training sets with binary images of rock samples from different regions. Due to the inherent randomness in EDSR training, the improvement rates in PSNR and SSIM did not follow a universal pattern with the upscale ratio. Comparing the datasets established by the bilinear and bicubic methods, the bilinear method-generated dataset led to a more substantial improvement in real low-resolution images, indicating that compared to bicubic-generated low-resolution images, those generated by the bilinear method are closer to real scanned low-resolution images. Fig. 13 also confirmed this result, showing that compared to different SEM images, using a bilinear method-generated image as the training set allowed the trained residual network to achieve a more significant improvement in real low-resolution images.

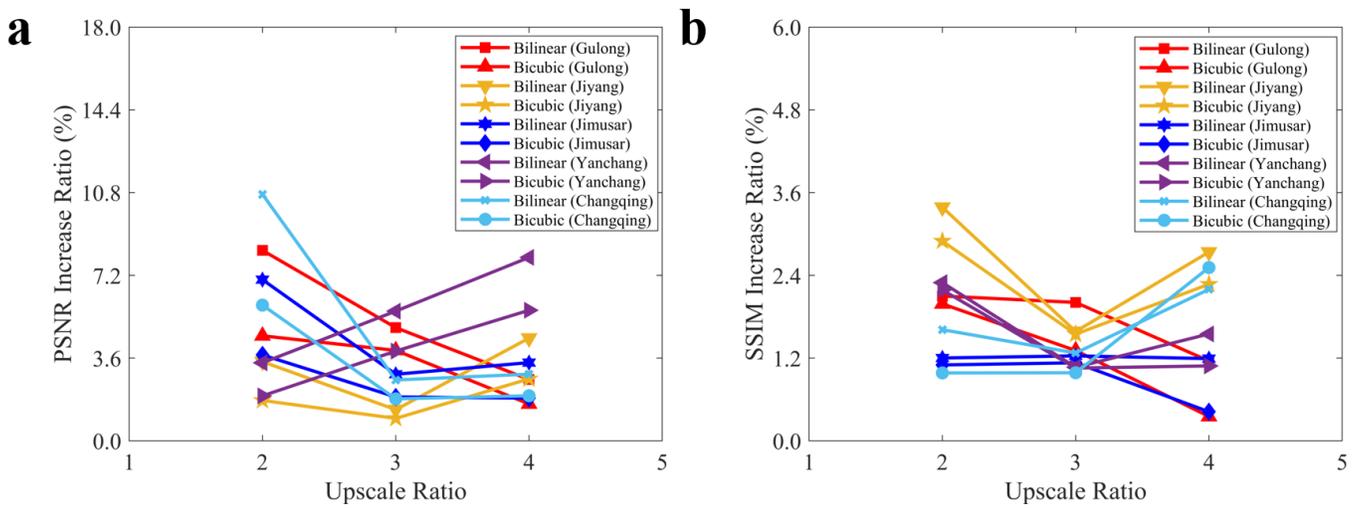

**Fig. 12—Images of rock samples from different regions: (a) PSNR increase ratio and (b) SSIM increase ratio of real low-resolution images by different EDSR training set.**

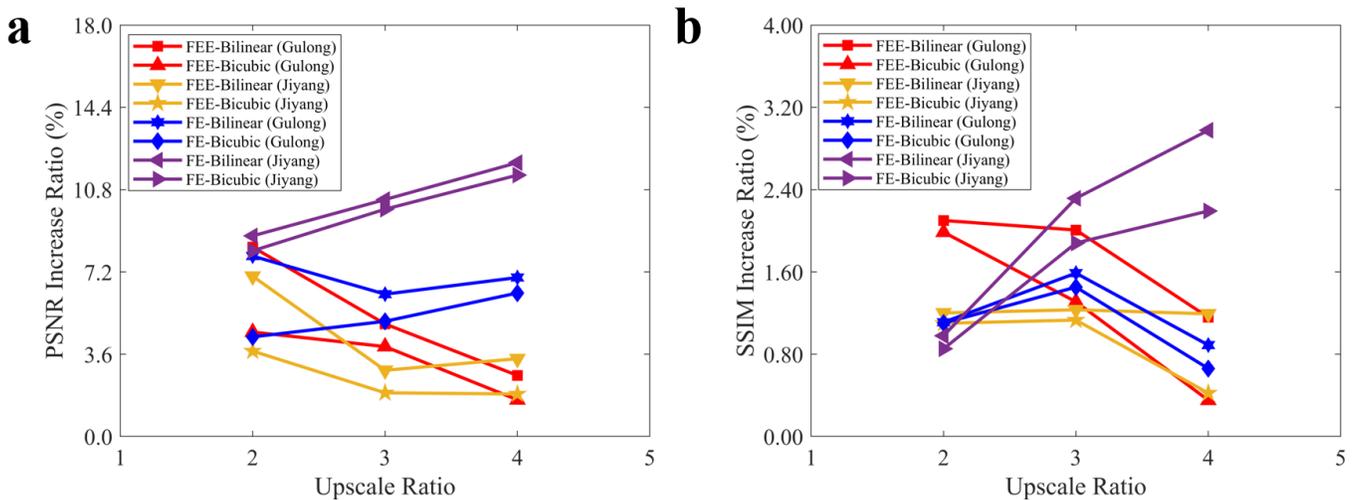

**Fig. 13—Images of different SEM scans: (a) PSNR increase ratio and (b) SSIM increase ratio of real low-resolution images by different EDSR training set.**

## Very Deep Super Resolution (VDSR)

To further validate our findings and demonstrate that they are not limited to a single super-resolution method, we trained the VDSR network using digital rock super-resolution training datasets generated by both the bilinear and bicubic methods. Fig. 14 displays the image reconstruction results for 2X, 4X, and 8X paired binary images, showing that the improvement in real low-resolution images using VDSR is less significant compared to EDSR. Therefore, it is also necessary to quantitatively characterize the extent of improvement that each training set provides to real low-resolution images.

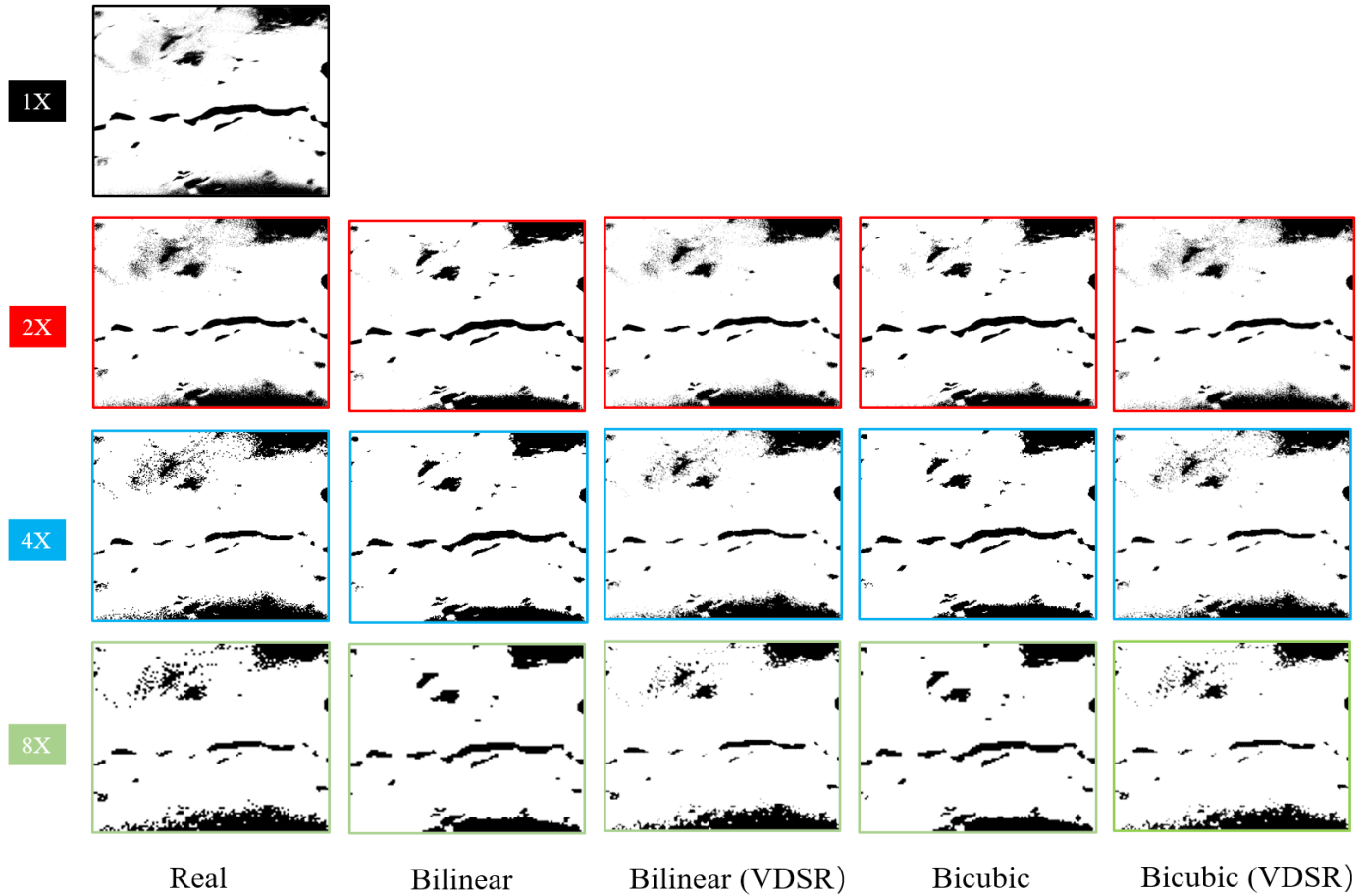

**Fig. 14—High-resolution images are reconstructed using residual networks trained with different VDSR training sets and real low-resolution images.**

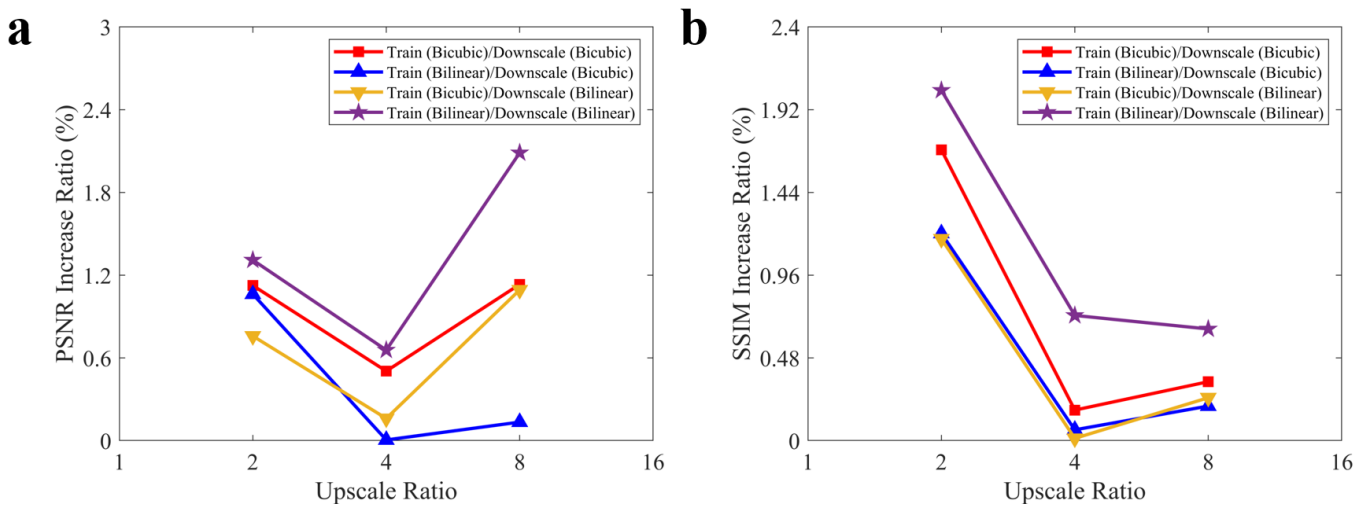

**Fig. 15—Comparison of (a) PSNR increase rates and (b) SSIM increase ratio of real low-resolution images by different VDSR training sets and image downscale method.**

Fig. 15 illustrates the percentage improvement in image quality metrics (PSNR and SSIM) achieved by residual networks trained with bilinear and bicubic datasets on real resolution images. Unlike EDSR, using the same method for generating paired images and resizing does not necessarily result in the largest improvement in image quality parameters, which we believe could be related to the randomness of VDSR training. Among the combinations of four training and downscaling methods, using the bilinear method for both generating paired images and resizing proved to be the most effective, achieving the highest increases in PSNR and SSIM for real low-resolution images. This underscores the superiority of the bilinear method in generating SEM digital rock paired training sets. Similar to EDSR, we compared grayscale images of rock samples from different regions and images scanned by different SEMs when testing VDSR with different training sets. Similar to EDSR, we compared grayscale images of rock samples from different regions and images scanned by different SEMs when testing VDSR with different training sets. The results for PSNR and SSIM are displayed in Figs. S-4 and S-5 in the Supplementary Materials, which also confirms that images generated by bilinear interpolation are the closest to the real scanned paired images.

## Discussion

### Grayscale and Binary Images

In this study, we compared grayscale images when selecting optimal upscale and downscale methods. Due to the limited size of the training set, we used binary images for method validation with EDSR and VDSR. This might raise questions about whether the principles applicable to grayscale images also apply to binary images. Theoretically, binary images converted from grayscale images that are closer to real images should also be closer to real binary images. We conducted tests using paired grayscale images as the EDSR training set, reconstructing high-resolution images from real low-resolution grayscale images (Fig. 16). The phenomenon observed is that the image quality improvement of real low-resolution grayscale images is minimal, and there is no significant difference between the residual networks trained on two different datasets. We hypothesize that this is due to the limited number of paired images in the training set, as only five sets of images were scanned for each type of rock sample. Additionally, the difference between the grayscale images at different resolutions is minimal, preventing the EDSR from effectively learning the detailed differences between images of different resolutions. Consequently, the trained residual network shows little improvement for real low-resolution images, and the differences between using bilinear and bicubic methods are not significant.

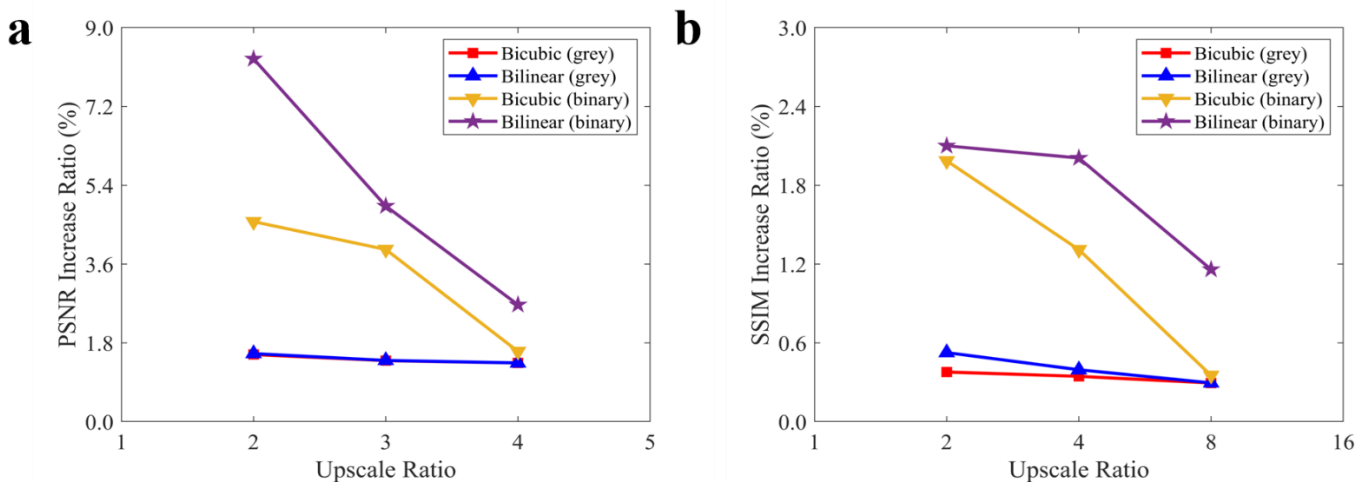

**Fig. 16—Using grayscale image and binary image to construct EDSR training set to compare (a) PSNR increase ratio and (b) SSIM increase ratio of real low-resolution image.**

To prove the effectiveness of our study, we compared images generated by different methods and, after converting them to binary images, evaluated their similarity to real binary images, please refer to Fig. S-6 of the Supplementary Materials for more details. The nearest method produced the largest disparity from real binary images, while the bilinear method produced binary images that were closest to real binary images. This conclusion aligns with direct comparisons of grayscale images. In future studies, we aim to expand the number of real paired images and use paired grayscale images as the super-resolution training set.

### Training Set Generation Methods

In sections 3.1 and 3.2, we tested six methods of image generation. As the bicubic method is the most commonly used for generating paired images, we mainly compared the bilinear and bicubic methods in sections 3.3 and 3.4, using these images as training sets for

EDSR and VDSR to compare the improvement in image quality metrics of real low-resolution images. In recent years, many researchers have also used the Lanczos2 and Lanczos3 methods to generate paired images for super-resolution training sets. We selected a set of images to test the effect of these six methods in generating training set paired images (Fig. 17). The results confirmed our previous findings: the bicubic, Lanczos2, and Lanczos3 methods had similar outcomes, but the bilinear method produced the best results for enhancing the quality of real low-resolution images.

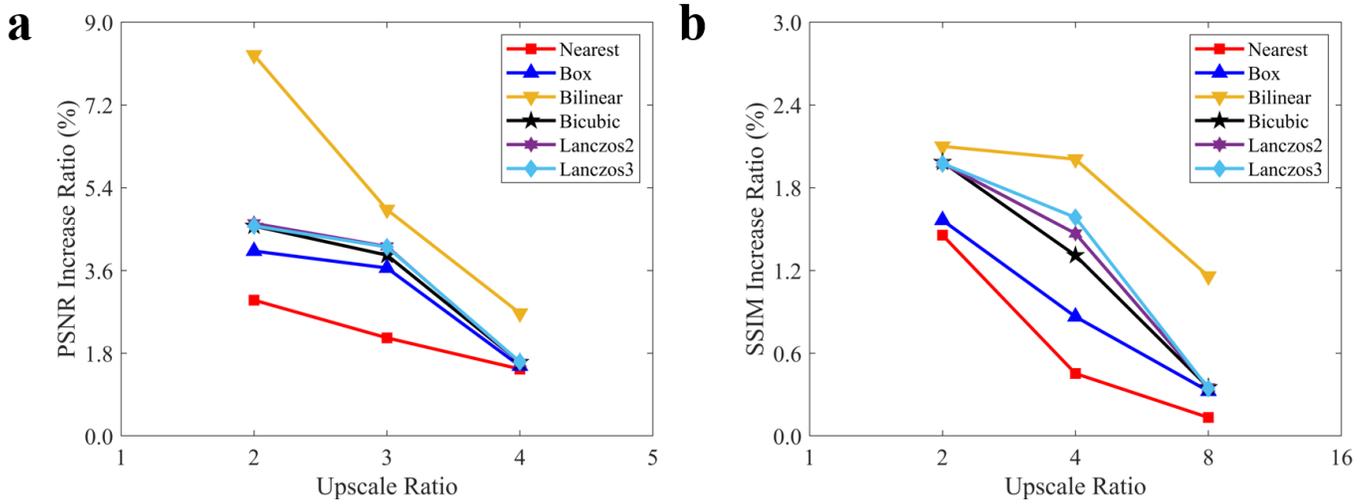

**Fig. 17—Comparison of (a) PSNR increase ratio and (b) SSIM increase ratio of real low-resolution images by constructing EDSR training sets with six methods.**

### Real Paired Image Training Set

We used different methods to generate paired images, which were then used as training sets to reconstruct high-resolution images using real low-resolution images and residual networks trained with different methods. By comparing the improvement rates in image quality metrics for real low-resolution images, we demonstrated that bilinear-generated paired images are closest to real scanned paired images. We can also prove this by reversing the process: using real scanned paired images as the super-resolution training set and reconstructing high-resolution images based on low-resolution images generated by different methods.

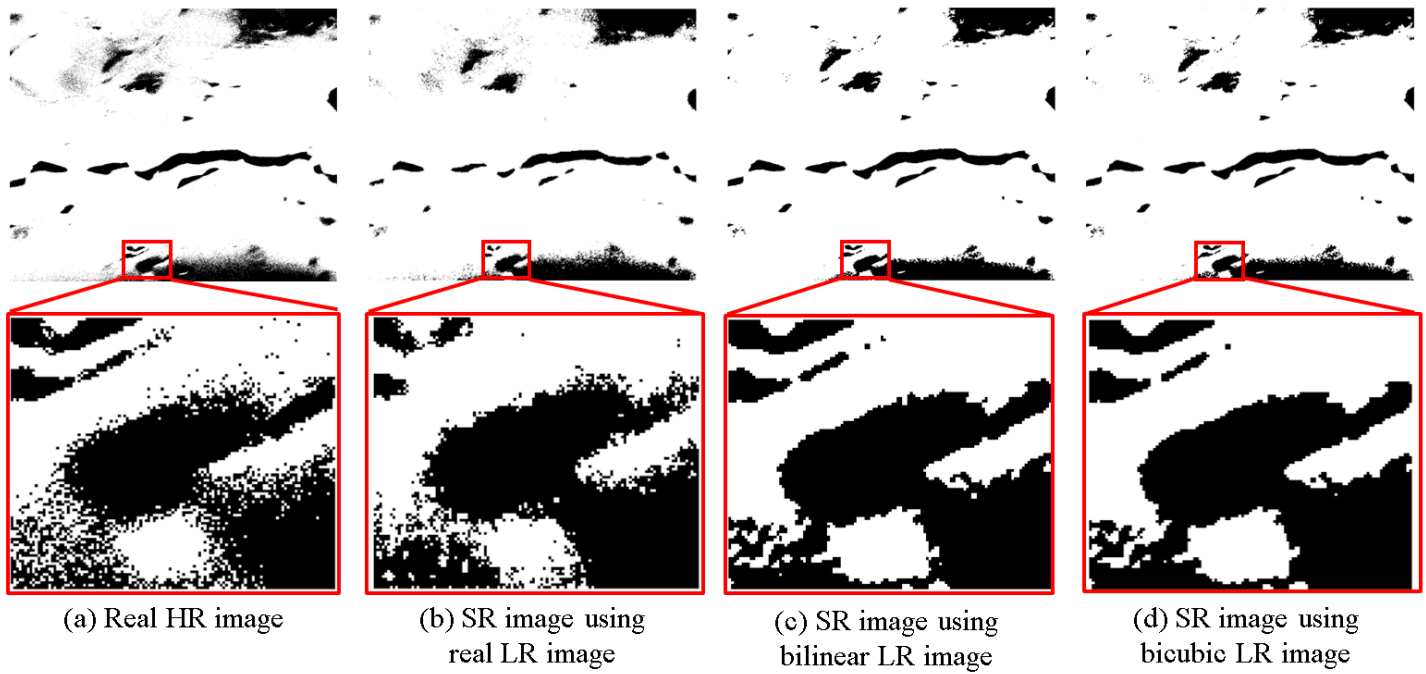

**Fig. 18—Detailed comparison of reconstructed high-resolution images from different low-resolution images.**

Fig. 18 presents high-resolution images and their detailed views: (a) real high-resolution image directly captured by SEM, and high-resolution images reconstructed using super-resolution from (b) real low-resolution images, (c) low-resolution images generated by the bilinear method, and (d) low-resolution images generated by the bicubic method. Since the training set consists of real paired images, the reconstruction using real low-resolution images yields the best results, with the reconstructed images closely resembling the real high-resolution images. The bilinear and bicubic methods show minimal improvement in image quality after reconstruction, making it challenging to directly compare the details of the two images. Therefore, quantitative analysis using PSNR and SSIM is necessary.

Fig. 18 shows the results, where we used EDSR to learn the detail differences between real high-resolution and low-resolution images, reconstructing high-resolution images using real low-resolution images, and low-resolution images generated by both bilinear and bicubic methods. As shown in Fig. 19, the residual network trained with real paired images achieved the greatest improvement in resolution for real low-resolution images. Compared to the bicubic method, the bilinear method resulted in a higher increase in PSNR and SSIM, which from another angle validates our findings.

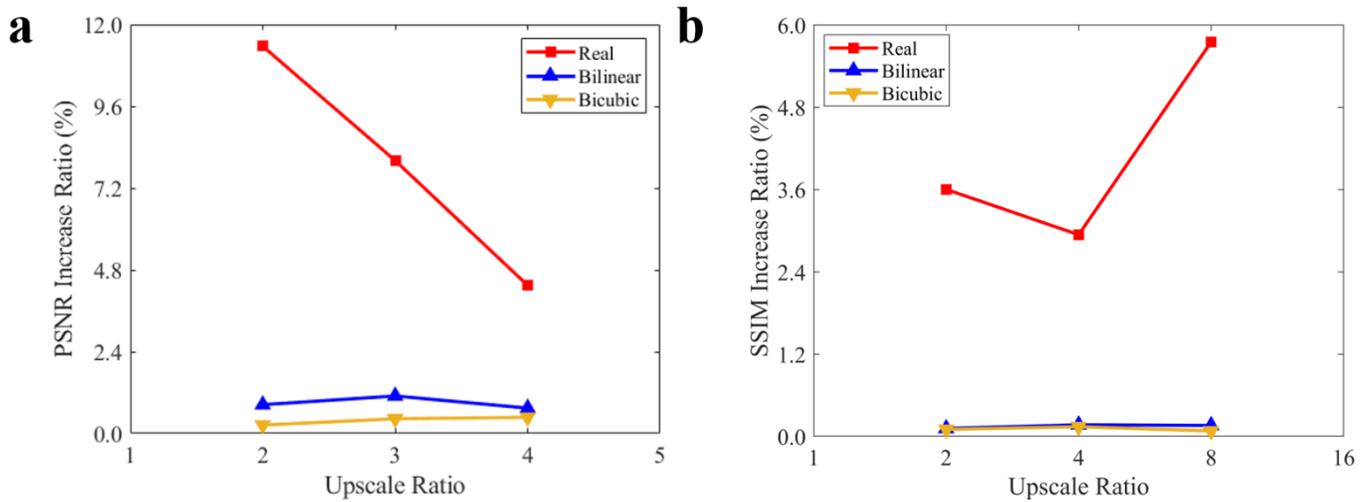

**Fig. 19**—(a) PSNR increase ratio and (b) SSIM increase ratio of the generated image by the training set composed of EDSR using real paired images.

### Image-forming Principle

We analyze the mechanism of image capture devices based on the phenomena observed in this study (Fig. 20a). Specifically, we first artificially generate a large 2D image representing the real object, and use six sampling methods (i.e., nearest, box, bilinear, bicubic, Lanczos2, and Lanczos3) to mimic/simulate the mechanism of image capture and hence obtain six sets of high-resolution (images A) and low-resolution (images B) paired images. For each set of paired images, we then upscale the high-resolution images (images A) using the six sampling methods and obtain the upscaled low-resolution images (images C). We compare images B and images C, and observe which upscaling method yields the closest results. The results show that for the images simulated using the box method, the low-resolution images upscaled using the bicubic or Lanczos methods are the closest to the simulated low-resolution images; whereas for the images simulated using the nearest method, the bilinear method provides the best results.

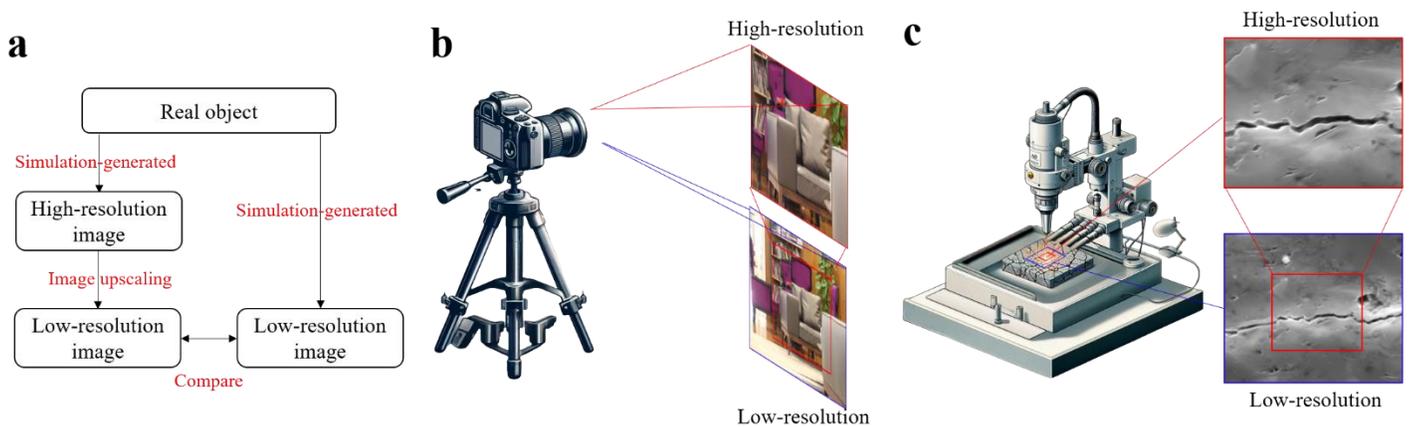

**Fig. 20**—(a) Flowchart for analyzing the image-forming principle, (b) schematic diagram of camera imaging principles and (c) schematic diagram of scanning electron microscope imaging principles.

Next, we test two type of image capture devices, i.e., SEM (Fig. 20b) and traditional cameras (Fig. 20c), from which we may infer the mechanism of image capture devices. For the SEM images, we upscale the real scanned high-resolution images using various methods and thus obtain the upscaled low-resolution images, which are compared to the real scanned low-resolution images. The results show that for all paired images captured by both the FE-SEM and FEE-SEM, including 2X, 4X, 8X and 16X, the nearest algorithm is the best method for image upscaling. Recalling the findings from the simulate results, we infer that the image capture mechanism of SEM is similar to the nearest method. This could be understood as the scanning mode of the probe of SEM, similar to Gaussian filtering. As for the traditional cameras, the real paired images in the datasets were captured using Nikon and Canon cameras (Cai et al., 2019; Wei et al., 2020). We employed the DRealSR dataset and Real-world SISR Dataset to assess which algorithm most accurately upscaled images to match the real paired images.

The DRealSR dataset is a comprehensive benchmark established to investigate the nuances of real-world super-resolution (SR) degradation. It employs zoom functions from five DSLR cameras—specifically, models from Canon, Sony, Nikon, Olympus, and Panasonic—to gather both low-resolution (LR) and high-resolution (HR) images. The dataset encompasses a variety of natural scenes, both indoors and outdoors, deliberately avoiding dynamic subjects such as advertising posters and plants. For each of the four scaling factors (i.e., 1x:4x), images are aligned using the SIFT method to ensure that each LR image corresponds accurately to its HR counterpart. To refine this alignment, images are segmented into patches, which are then subjected to iterative registration and brightness matching to maintain fidelity between the LR and HR images. For efficient model training, the images, originally sized between 3888×5184 and 4000 6000 pixels, are cropped into patches measuring 380×380, 272×272, and 192×192 for scaling factors of 2x:4x, respectively.

The Real-world SISR dataset is developed using two advanced full-frame DSLR cameras: the Canon 5D Mark III and the Nikon D810, featuring resolutions of 5760×3840 and 7360×912 pixels, respectively. Each camera was outfitted with a 24-105mm f/4.0 zoom lens, enabling coverage of scaling factors commonly utilized in super-resolution research, namely 2x, 3x, and 4x. For each captured scene, photographs were taken at four distinct focal lengths: 105mm, 50mm, 35mm, and 28mm. The images obtained at the 105mm focal length form the basis for the high-resolution (HR) ground truth, while those captured at the lesser focal lengths are used to construct the low-resolution (LR) variants. A focal length of 28mm was preferred over 24mm to minimize lens distortion, which is more challenging to correct in post-processing and could degrade the quality of aligned image pairs.

We upscale the real filmed high-resolution images using various methods and thus obtain the upscaled low-resolution images, which are compared to the real filmed low-resolution images. The results show that for all paired images captured by the two cameras, including 2X, 3X and 4X, the Lanczos algorithm is the best method for image upscaling. Recalling the findings from the simulate results, we suspect that the image capture mechanism of traditional cameras is more like the box method. This can be understood as the sensors in cameras capture the intensity of light, and convert it into electrical signals. These electrical signals are then processed and stored, forming a digital image. In cameras, there are two common image capture mechanisms: CCD (Charge-Coupled Device) and CMOS (Complementary Metal-Oxide-Semiconductor). CCD image sensors convert light into electrical charges, which are then transferred to a register and read out. This method can provide high-quality images, but it requires more power and has a slower readout speed. CMOS image sensors, on the other hand, use integrated circuits to convert light into electrical signals, which are then read out. This method can provide fast readout speeds and low power consumption, but it may introduce more noise and inferior image quality. Moreover, some cameras may employ additional techniques to enhance image quality, such as image interpolation, demosaicing, and noise reduction. Understanding the different image capture mechanisms and technologies can help us appreciate the complexity and beauty of image capture systems, while further analysis is required, which is beyond the scope of this study.

**Conclusion**

This research aims to identify the image generation method that most closely matches real scanned SEM paired images for use in digital rock super-resolution training datasets. Super-resolution technology primarily involves two steps: network training and image reconstruction. Initially, convolutional neural networks or adversarial networks learn the detail differences between images of different resolutions, followed by using low-resolution images and trained networks to reconstruct high-resolution images. The training step requires paired images of different resolutions, but due to high scanning costs and difficulties in image pairing, real scanned paired images are challenging to obtain. Researchers in the super-resolution field typically use bicubic and Lanczos algorithms to generate paired images, which are well-suited for images captured by cameras. However, because the scanning principles of SEM equipment vary significantly from camera photography, these algorithms may not be directly applicable to the digital rock super-resolution domain.

In this study, we obtained grayscale images of shale samples from five regions using the Hitachi SU8010 Field-Emission Scanning Electron Microscope (FE-SEM) and a Quanta 200F Field-Emission Environmental Scanning Electron Microscope (FEE-SEM), scanning five images per sample at magnifications of 1X, 2X, 4X, 8X, and 16X. We used these real scanned paired images as a reference to select the most appropriate image generation method. We tested six algorithms: nearest, box, bilinear, Lanczos2, and Lanczos3, and used PSNR and SSIM as image quality metrics to quantitatively compare the images generated by these methods, ultimately selecting the bilinear algorithm. We used paired images generated by this method and the commonly used bicubic method as super-resolution training sets, employing EDSR and VDSR to compare the improvement rates in image quality metrics for real low-resolution images. The bilinear method proved to generate images that more closely resemble SEM scanned images. Our main contributions are twofold. First, we propose the use of the bilinear algorithm to generate SEM super-resolution training set paired images, which is more suitable for the digital rock super-resolution domain compared to the commonly used bicubic and Lanczos algorithms. Second, we obtained a variety of SEM high-resolution and low-resolution paired images using two types of scanning microscopes. By scanning different resolution paired images of shale samples from five regions, we demonstrated the universality of our conclusions.

This study proposed an image generation method suitable for the digital rock domain, capable of constructing digital rock super-resolution datasets that closely mimic real paired images. Different from the common knowledge or practice, the proposed approach provides new insights for establishing digital rock super-resolution datasets. Possible extensions such as generating paired CT images in 3D (i.e., the same area in both high-resolution and low-resolution) are currently being investigated.

**Acknowledgements**

The authors would like to acknowledge the support provided by the National Natural Science Foundation of China (No. 52374017), National Key Laboratory of Petroleum Resources and Engineering (PRE/indep-1-2303) and Science Foundation of China University of Petroleum, Beijing (No. 2462022QNXZ002).

**References**

Bai, Y., Berezovsky, V., & Popov, V., (2020). Super Resolution for Digital Rock Core Images via FSRCNN. In Proceedings of the 2020 4th High Performance Computing and Cluster Technologies Conference & 2020 3rd International Conference on Big Data and Artificial Intelligence (HPCCT & BDAI '20). Association for Computing Machinery, New York, NY, USA, 78–81.

Bizhani, M., Ardakani, O. H., & Little, E. (2022). Reconstructing high fidelity digital rock images using deep convolutional neural networks. Scientific Reports, 12(1), 1-14.

Blunt, M. J., Bijeljic, B., Dong, H.et al. (2013). Pore-scale imaging and modelling. Advances in Water resources, 51, 197-216.

Cai, J., Zeng, H., Yong, H.et al. (2019). Toward real-world single image super-resolution: A new benchmark and a new model. In Proceedings of the IEEE/CVF international conference on computer vision (pp. 3086-3095).

Chang, H., Yeung, D. Y., & Xiong, Y. (2004, June). Super-resolution through neighbor embedding. In Proceedings of the 2004 IEEE Computer Society Conference on Computer Vision and Pattern Recognition, 2004. CVPR 2004. (Vol. 1, pp. I-I). IEEE.

Chaves, J. M., & Moreno, R. B. (2021). Low-and high-resolution X-ray tomography helping on petrophysics and flow-behavior modeling. Spe Journal, 26(01), 206-219.

Chen Honggang, He Xiaohai, Teng Qizhi.et al. (2020). Super-resolution of real-world rock microcomputed tomography images using cycle-consistent generative adversarial networks. Physical review. E (2-1),023305.

Dong, C., Loy, C. C., He, K.et al. (2014). Learning a deep convolutional network for image super-resolution. In Computer Vision–ECCV 2014: 13th European Conference, Zurich, Switzerland, September 6-12, 2014, Proceedings, Part IV 13 (pp. 184-199). Springer International Publishing.

Duchon, C. E. (1979). Lanczos filtering in one and two dimensions. Journal of Applied Meteorology and Climatology, 18(8), 1016-1022.

Flannery, B. P., Deckman, H. W., Roberge, W. G.et al. (1987). Three-dimensional X-ray microtomography. Science, 237(4821), 1439-1444.

Freeman, W. T., Jones, T. R., & Pasztor, E. C. (2002). Example-based super-resolution. IEEE Computer graphics and Applications, 22(2), 56-65.

Fang, H. H., Wang, Z. F., Sang, S. X., & Huang, Y. H. (2023). Numerical analysis of matrix swelling and its effect on microstructure of digital coal and its associated permeability during CO2-ECBM process based on X-ray CT data. Petroleum Science, 20(1), 87-101.

Hu, M., & Tan, J. (2006). Adaptive osculatory rational interpolation for image processing. Journal of Computational and Applied Mathematics, 195(1-2), 46-53.

Iglewicz, B., & Hoaglin, D. (1993). Volume 16: How to detect andhandle outliers. In E. F. Mykytka (Ed.), The ASQC basic ref-erences in quality control: Statistical techniques (pp. 1–77). American Society for Quality Control Statistics Division.

Kim, J., Lee, J. K., & Lee, K. M. (2016). Accurate image super-resolution using very deep convolutional networks. In Proceedings of the IEEE conference on computer vision and pattern recognition (pp. 1646-1654).

Kang, J., Li, N. Y., Zhao, L. Q., Xiong, G., Wang, D. C., & Xiong, Y., et al. (2022). Construction of complex digital rock physics based on full convolution network. Petroleum Science, 19(2), 651-662.

Krakowska, P., Dohnalik, M., Jarzyna, J.et al. (2016). Computed X-ray microtomography as the useful tool in petrophysics: A case study of tight carbonates Modryn formation from Poland. Journal of Natural Gas Science and Engineering, 31, 67-75.

Ledig, C., Theis, L., Huszár, F.et al. (2017). Photo-realistic single image super-resolution using a generative adversarial network. In Proceedings of the IEEE conference on computer vision and pattern recognition (pp. 4681-4690).

Lewis, J. P. (1995, May). Fast normalized cross-correlation. In Vision interface (Vol. 10, No. 1, pp. 120-123).

Li, Z., Teng, Q., He, X.et al. (2017). Sparse representation-based volumetric super-resolution algorithm for 3D CT images of reservoir rocks. Journal of Applied Geophysics, 144, 69-77.

Liang, Y., Wang, S., Feng, Q., Zhang, M., Cao, X., & Wang, X. (2024). Ultrahigh-resolution reconstruction of shale digital rocks from FIB-SEM images using deep learning. SPE J., 29(3), 1434-1450.

Liao, Q., Xue, L., Wang, B.et al. (2022). A new upscaling method for microscopic fluid flow based on digital rocks. Advances in Geo-Energy Research, 6(4), 357-358.

Lim, B., Son, S., Kim, H.et al. (2017). Enhanced Deep Residual Networks for Single Image Super-Resolution. 2017 IEEE Conference on Computer Vision and Pattern Recognition Workshops (CVPRW), 1132-1140.

Lv, X., Ming, D., Chen, Y.et al. (2019). Very high resolution remote sensing image classification with SEEDS-CNN and scale effect analysis for superpixel CNN classification. International Journal of Remote Sensing, 40(2), 506-531.

Shan, L., Bai, X., Liu, C.et al. (2022). Super-resolution reconstruction of digital rock CT images based on residual attention mechanism. Advances in Geo-Energy Research, 6(2), 157-168.

Starnoni, M., Pokrajac, D., & Neilson, J. E. (2017). Computation of fluid flow and pore-space properties estimation on micro-CT images of rock samples. Computers & Geosciences, 106, 118-129.

Vb, S. K. (2020). Perceptual image super resolution using deep learning and super resolution convolution neural networks (SRCNN). Intelligent Systems and Computer Technology, 37(3).

Wang, Y., Rahman, S. S., & Arns, C. H. (2018). Super resolution reconstruction of µ-CT image of rock sample using neighbour embedding algori0thm. Physica A: Statistical Mechanics and its Applications, 493, 177-188.

Wei, P., Xie, Z., Lu, H.et al. (2020). Component divide-and-conquer for real-world image super-resolution. In Computer Vision–ECCV 2020: 16th European Conference, Glasgow, UK, August 23–28, 2020, Proceedings, Part VIII 16 (pp. 101-117). Springer International Publishing.

Xing, Z., Yao, J., & Liu, L. (2023). Digital rock resolution enhancement and detail recovery with multi attention neural network. Geoenergy Science and Engineering. Volume 227, 211920.

Yang, Y., Horne, R. N., Cai, J.et al. (2023). Recent advances on fluid flow in porous media using digital core analysis technology. Advances in Geo-Energy Research, 9(2).

Yang, Y., Zhou, Y., Blunt, M. J.et al. (2021). Advances in multiscale numerical and experimental approaches for multiphysics problems in porous media. Advances in Geo-Energy Research, vol. 5, no. 3, pp. 233-238.

Yu, J., Fan, Y., Yang, J.et al. (2018). Wide activation for efficient and accurate image super-resolution. arxiv preprint arxiv:1808.08718.

Yu, F., Han, R., & Xu, K. (2024). Countercurrent Imbibition in Porous Media with Dense and Parallel Microfractures: Numerical and Analytical Study. SPE Journal, 1-20.

Yuan, B., Li, H., & Du, Q. (2023). Enhancing identification of digital rock images using super-resolution deep neural network. Geoenergy Science and Engineering, 229, 212130.

Zhang, X., Chen, Q., Ng, R.et al. (2019). Zoom to learn, learn to zoom. In Proceedings of the IEEE/CVF Conference on Computer Vision and Pattern Recognition (pp. 3762-3770).

Zhao, B., Saxena, N., Hofmann, R.et al.(2022). Enhancing resolution of micro-CT images of reservoir rocks using super resolution. Comput. Geosci., 170, 105265.